\documentclass[aps,nofootinbib,twocolumn,superscriptaddress,groupedaddress]{revtex4-1}
\usepackage{adjustbox}
\usepackage{dcolumn}   
\usepackage{bm,amsmath,amssymb}
\usepackage{graphicx}
\usepackage{subfigure}
\usepackage{soul}
\usepackage[normalem]{ulem}
\usepackage[dvipsnames]{xcolor}

\sethlcolor{yellow}  
\usepackage{mathtools}
\usepackage{mathtools}
\usepackage{amsthm}
\usepackage{braket}
\usepackage[english]{babel}
\usepackage[utf8x]{inputenc}
\usepackage[colorlinks=true,linkcolor=blue]{hyperref}
\usepackage{color}

\definecolor{Ora}{cmyk}{0, 0.6, 0.8, 0}
\makeatletter
\renewcommand{\p@subsection}{}
\renewcommand{\p@subsubsection}{}
\newcommand*\bigcdot{\mathpalette\bigcdot@{.5}}
\newcommand*\bigcdot@[2]{\mathbin{\vcenter{\hbox{\scalebox{#2}{$\m@th#1\bullet$}}}}}
\makeatother

\def\ket#1{|#1\rangle}
\def\bra#1{\langle#1|}
\def\scal#1#2{\langle#1|#2\rangle}
\def\matr#1#2#3{\langle#1|#2|#3\rangle}

\def\ave#1{\langle#1\rangle}

\def\tr{\text{tr}}

\def\smallunderbrace#1{\mathop{\vtop{\m@th\ialign{##\crcr
				$\hfil\displaystyle{#1}\hfil$\crcr
				\noalign{\kern3\p@\nointerlineskip}%
				\tiny\upbracefill\crcr\noalign{\kern3\p@}}}}\limits}

\begin{document}
	
	
	\title{Driving of an open quantum system at finite temperature across first- and second-order quantum phase transitions}
	
	\author{Felipe Matus}
	\email{matus@ipnp.mff.cuni.cz}
	\affiliation{Institute of Particle and Nuclear Physics, Faculty of Mathematics and Physics, Charles University,
		V Hole{\v s}ovi{\v c}k{\' a}ch 2, 180 00 Prague, Czechia}
	\author{Pavel Cejnar}
	\email{cejnar@ipnp.mff.cuni.cz}
	\affiliation{Institute of Particle and Nuclear Physics, Faculty of Mathematics and Physics, Charles University,
		V Hole{\v s}ovi{\v c}k{\' a}ch 2, 180 00 Prague, Czechia}
  
	\date{\today}
	
	\begin{abstract}
    An open fully connected system of qubits at nonzero temperature is driven within a finite time interval along various paths in the space of its control parameters.
    The driving leads across finite-size precursors of first- and second-order quantum phase transition from factorized to entangled ground-state phases, aiming at the preparation of the complex ground state of the system at the final parameter point with maximal fidelity.
    During the drive, the system is coupled to a heat bath at a constant temperature, the dynamics being determined in a nonperturbative way by the method of Hierarchical Equations of Motion. 
    It is shown that the presence of the heat bath in combination with specific patterns of avoided crossings affecting the ground and excited states in the parameter region around the quantum phase transition may considerably improve the fidelity of preparation of the target ground state. 
	\end{abstract}
	
	\maketitle
	
\section{INTRODUCTION}
\label{sec:intro}

One of the aims of arising quantum information science is to develop efficient and reliable techniques for preparing general correlated states of complex quantum systems~\cite{Grov00,Ples11}. 
A common strategy is based on driving a suitable system involving complex interactions among its elementary constituents along a selected path in the space of control parameters~\cite{Farh00,Ahar07a,Ahar07b,Alba18}.
The path starts at an initial point corresponding to an uncorrelated, easy-to-prepare ground state, and terminates at a final point, where the ground state contains the desired complex quantum correlations.
Reproducing the target ground state with high fidelity and in moderate time is a task that has direct applications in quantum control, quantum computation and information processing~\cite{Niel02}. 

In the last decades different strategies have been proposed to accomplish this task.
One can mention, among the most striking approaches, the adiabatic, transitionless, geometric, and decoherence-assisted types of driving.       
The adiabatic driving represents a direct application of the quantum adiabatic theorem, which states that for a sufficiently slow variation of control parameters, the system follows the instantaneous (adiabatic) ground state in the parameter space~\cite{Kato50,Jans07,Elga12}.
The transitionless (also called counter-diabatic) driving emulates the adiabatic evolution along a given parameter path with the aid of additional terms in the Hamiltonian (which usually are non-local) that compensate non-adiabatic effects of the finite-speed driving~\cite{Demi03,Berr09,Camp13,Guer19}. 
The geometric driving makes use of the geodesic path between the initial and final parameter points with respect to the Provost-Vallee metric on the ground-state manifold in attempt to maximize the overlap of the evolving state with the adiabatic ground state~\cite{Prov80,Anan90,Miya01,Reza09,Kolo17,Buko19}.
Finally, the decoherence-assisted driving increases the fidelity of the final-state preparation by repeated measurement-like interactions of the driven system with a suitable ancilla~\cite{Chil02,Roa06,Haco18,Cejn23}. 

Some of these driving strategies were recently tested by us in systems composed of a single or multiple interacting qubits~\cite{Cejn23,Matu23a,Matu23b}.
The multi-qubit environment was implemented through a model from the Lipkin-Meshkov-Glick (LMG) family~\cite{Lipk65}, which is numerically treatable and available to experiments, but simultaneously shows complex phenomena like ground-state and excited-state quantum phase transitions~\cite{Gilm78,Orus08,Zibo10,Puri17,Cerv21}.
The ground-state quantum phase transitions (QPTs), as sudden changes (non-analytic in the limit of infinite system size) of the ground-state energy and wave function with Hamiltonian control parameters~\cite{Sach99}, are of a straightforward importance for the driving problem if the initial and target states belong to different ground-state phases.
Crossing of the QPT point usually sets the most stringent bounds on the total driving times and/or the final fidelity achieved~\cite{Lato04,Zure05,Schu06}.
On the other hand, excited-state quantum phase transitions (ESQPTs), which represent an extension of the QPT to the excited domain~\cite{Cejn06,Ribe07,Capr08,Cejn08,Pere09,Sant16,Kopy17,Cejn21}, become relevant as soon as quantal or thermal fluctuations create nonvanishing populations of states in the ESQPT domain during the driving process.
This typically applies to driving through a wider QPT region, where ESQPT structures appear at low energies and have considerable effect on dynamics of the ground-state occupation.

Our previous studies~\cite{Cejn23,Matu23a,Matu23b} were focused on fully coherent and decoherence-assisted types of driving in strictly isolated systems only.
This means that the system was initiated at exactly zero temperature and excited states of the intermediate Hamiltonians were populated solely by quantum transitions induced by non-adiabaticity of the driving.
However, in realistic situations, the driven system is likely to interact---at least to a limited extent---with its surrounding environment.
This leads to thermal population of excited states during the course of the driving process.
If the driving leads across the QPT and ESQPT parameter regions, thermal noise can excite the system due to reduced energy gaps and relax it afterward. Thus, the impact of temperature on the final fidelity may become relatively strong~\cite{Dick13,Alba17}
To analyze such effects is the main purpose of the present paper.  

Open systems~\cite{Breu02,Weis12,Schl07} are most commonly studied within the Lindblad formalism, i.e., assuming that the system is weakly coupled to a Markovian reservoir and considering the Born and rotating wave approximations~\cite{Alba12,Yama17,Dann18}. 
Although the equations obtained in this way are numerically simple to solve, their validity is subject to too stringent conditions and the results are not guaranteed to be accurate~\cite{Ishi05,Dijk12,Iles14,Iles16,Abbo20,Anto23_1,Anto23_2}.
In this work, we use the method of Hierarchical Equations of Motion (HEOM)~\cite{Tani89,Tani06,Fruch16,Tani20,Lamb23}, which is based on a numerically exact solution of quantum dynamics through the influence functional formalism~\cite{Feyn63,Cald81,Cald83}. 
This approach is applicable to a large span of driving times and allows for narrow energy gaps.
The method has already been used to study quantum dissipative dynamics under time-dependent driving fields~\cite{Xu09}, entanglement dynamics~\cite{Ma12} as well as more complex dynamics~\cite{Ishi09,Chen09,Stru11,Lamb19}.
	
The structure of this paper is as follows:
In Sec.\,\ref{sec:model} we outline the model, along with its quantum critical properties, and describe the quantities that characterize the environment and its interaction with the system. 
In Sec.\,\ref{sec:driving_procedure} we introduce the driving procedures used to evolve the system and briefly describe the HEOM method and its parameters. 
In Sec.\,\ref{sec:results_discussion} we present and discuss the main results of our numerical calculations. 
In Sec.\,\ref{CONCLUSION}, we give a brief summary and conclusion.
	
\section{Model}
\label{sec:model}

We consider an interacting system of qubits (hereafter denoted by S) coupled to an external environment (denoted by E). 
The total Hamiltonian is decomposed into three parts:
\begin{equation}\label{Total_Hamiltonian}
	\hat{H} = \hat{H}_{\text{S}}\bigl(\bold{\Lambda}\bigr)  +  \hat{H}_{\text{E}} + \hat{H}_{\text{I}} ,
\end{equation}
where $\hat{H}_{\text{S}}\bigl(\bold{\Lambda}\bigr)$ is the Hamiltonian of qubits, with $\bold{\Lambda}$ denoting the set of control parameters that will be later subject to the driving procedure, $\hat{H}_{\text{E}}$ is the Hamiltonian of the environment (also called the bath), and $\hat{H}_{\text{I}}$ stands for the system-environment interaction. 
The forms of these Hamiltonians are given below.

Note that in this paper, the overall energy scale is given by an implicit constant $\varepsilon$, whose choice is arbitrary. 
So all components of the Hamiltonian~\eqref{Total_Hamiltonian} and in general all energies~$E$ are considered dimensionless, expressed in units of~$\varepsilon$.
Since we additionally set ${\hbar = k_{\text{B}} =1}$, the temperature~$T$ is given in units of $\varepsilon$ and the time $t$ in units of~$1/\varepsilon$.
	
\subsection{Qubit system}

The system S consists of $N >1$ mutually interacting qubits, equivalent to spin-$\frac{1}{2}$ particles, whose dynamics is described within the LMG framework~\cite{Lipk65}. 
The Hamiltonian is written in terms of collective spin (quasispin) operators
\begin{equation}
	\hat{J}_{\alpha} = \frac{1}{2} \sum_{i = 1}^{N}\hat{\sigma}_{\alpha}^{(i)}, \quad \alpha = x, y, z,
\end{equation}
where $\hat{\sigma}_{\alpha}^{(i)}$ are Pauli matrices acting on the $i$th-qubit space. 
The operators $\hat{J}_{\alpha}$ satisfy the usual SU(2) commutation relations. 
	
We use the LMG Hamiltonian used in our previous studies~\cite{Cejn23,Matu23a},
\begin{eqnarray} \label{H_S}
    \hat{H}_{\text{S}}(\bold{\Lambda})= \hat{J}_z-\frac{1}{N}\biggl[\lambda\hat{J}_x^2&+&\chi\left\{\hat{J}_x,\hat{J}_z\!+\!\tfrac{N}{2}\right\}
    \nonumber\\
    &+&\chi^2\bigl(\hat{J}_z\!+\!\tfrac{N}{2}\bigr)^2\biggr],
    \label{H_Lipkin}
\end{eqnarray}
where ${\{A,B\}=AB+BA}$.
The first term $\hat{J}_z$ of this Hamiltonian describes a system of noninteracting qubits, while the terms in square brackets represent interactions connecting any qubit with all the other qubits of the system.
The dimensionless interaction strengths $\lambda$ and $\chi$, jointly denoted as $\bold{\Lambda}\equiv\{\Lambda^{\mu}\}_{\mu=1,2}$, are considered as the system control parameters.
These will be subject (see Sec.\,\ref{dripro} below) to an externally driven time variation, $\bold{\Lambda}(t) = (\lambda(t),\chi(t))$, which will make the system Hamiltonian time-dependent:  $\hat{H}_{\text{S}}(t)=\hat{H}_{\text{S}}(\bold{\Lambda}(t))$.
As seen in Eq.\,\eqref{H_S}, the above-introduced energy constant~$\varepsilon$, in units of which the Hamiltonian is expressed, coincides with the energy difference between the up and down qubit states in absence of interaction (${\lambda=\chi=0}$).
	
The total quasispin ${\hat{J}^2 = \hat{J}_{x}^2 + \hat{J}_{y}^2 +\hat{J}_{z}^2}$ is conserved by $\hat{H}_{\text{S}}(\bold{\Lambda})$, so the full $2^{N}$-Hilbert space of qubits splits into a sum of $(2j+1)$-dimensional subspaces (almost all of them appearing in numerous replicas) with different values of the quantum number $j$~\cite{Cejn16}. 
We restrict here to the unique $(N+1)$-dimensional subspace with $j = \frac{N}{2}$, which is fully symmetric under the exchange of qubits, and assume that this subspace is invariant under the full Hamiltonian \eqref{Total_Hamiltonian} since also the interaction term $\hat{H}_{\text{I}}$ conserves the exchange symmetry.
The energy spectrum of $\hat{H}_{\text{S}}(\bold{\Lambda})$ is expressed in an ordered form: $E_0(\bold{\Lambda})\leq E_1(\bold{\Lambda})\leq \dots\leq E_N(\bold{\Lambda})$.

For $\chi = 0$, another quantity, which is conserved by the system Hamiltonian, is the parity
\begin{equation}\label{parity}
		\hat{\mathcal{P}} = (-1)^{\hat{J}_z-\frac{N}{2}}.
\end{equation}
However, in the following we will assume that this symmetry is not necessarily preserved by the interaction Hamiltonian $\hat{H}_{\text{I}}$.

The ground state of Hamiltonian \eqref{H_Lipkin} exhibits QPTs of first and second order.
These can be described in terms of two order parameters associated with the ground-state expectation values ${\braket{\hat{J}_z + \frac{N}{2}}}$ and $\braket{\hat{J}_x}$.
For ${N\to\infty}$, the plane of control parameters ${\lambda\times\chi}$ splits in three ground-state phases:
phase I with ${\braket{\hat{J}_z + \frac{N}{2}}=\braket{\hat{J}_x}=0}$ in the domain ${\lambda < \lambda_{\rm c}(\chi)}$, where
\begin{equation}\label{critical}
    \lambda_{\rm c}(\chi)= 1- \frac{\chi^2}{1-\chi^2},
\end{equation}
phase II with ${\braket{\hat{J}_z + \frac{N}{2}}>0}$ and ${\braket{\hat{J}_x}>0}$ in the domain ${\lambda>\lambda_{\rm c}(\chi)}$, ${\chi>0}$, and phase III with ${\braket{\hat{J}_z + \frac{N}{2}}>0}$ and ${\braket{\hat{J}_x}<0}$ in the domain ${\lambda>\lambda_{\rm c}(\chi)}$, ${\chi<0}$.
The ${N\to\infty}$ form of the ground state in phase I is fully factorized, corresponding to separated qubits in the down spin projection states, whereas in phases II and III all qubits in the ground-state wave function are mutually entangled.
Since phases II and III are symmetric with respect to the ${\chi\leftrightarrow-\chi}$ inversion, we consider here only the case ${\chi\geq 0}$ with phases I and II.

\begin{figure}[tp]
    \includegraphics[width=1.0\linewidth]{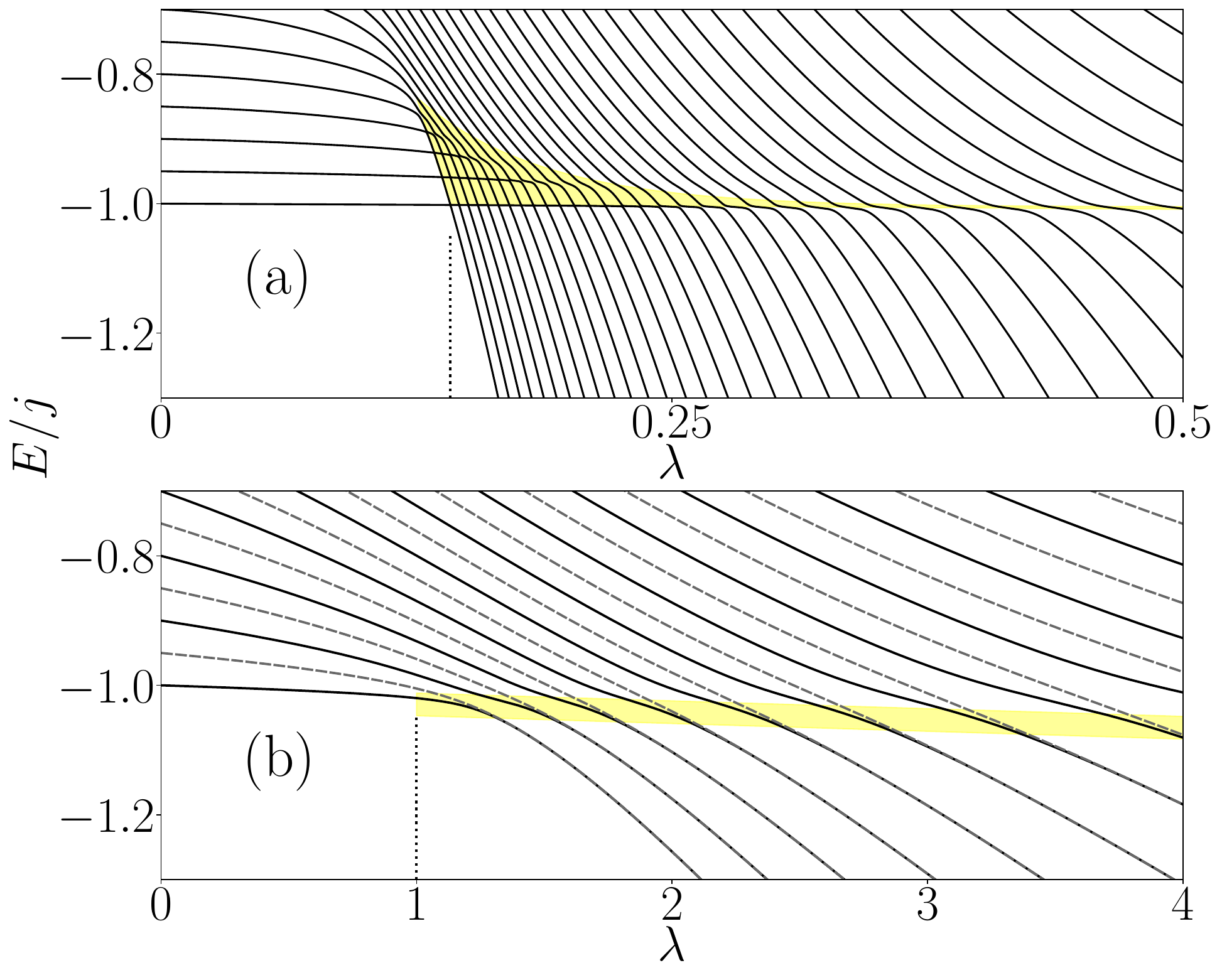}
    \caption{Energy spectra of the LMG Hamiltonian \eqref{H_Lipkin} for two cuts through the ${\lambda\times\chi}$ parameter plane: (a) ${\lambda\in[0,0.5]}$, ${\chi=4.8\lambda}$ and (b) ${\lambda\in[0,4]}$, ${\chi=0}$. We set ${N=40}$. The vertical dotted line segments demarcate ${N\to\infty}$ critical points $\lambda_{\rm c}(\chi)$ of (a) the first-order QPT and (b) the second-order QPT. In panel (b) the positive and negative parity levels are drawn by full and dashed curves, respectively, visualizing the spontaneous parity breaking at the second-order QPT. All level crossings are avoided. Their patterns (highlighted by the background color) around both ground-state QPTs represent precursors of ESQPTs.
    }
    \label{fig:fig1}
\end{figure}

The QPTs between these phases are of the first order for ${\chi\neq 0}$ and of the second order for ${\chi=0}$~\cite{Ribe07}.
In the first-order QPT, the order parameters change discontinuously, while the ground-state energy $E_0(\bold{\Lambda})$ shows a discontinuity of the first derivative and the ground-state energy gap $\Delta_{10}(\bold{\Lambda})={E_1(\bold{\Lambda})-E_0(\bold{\Lambda})}$ vanishes exponentially with ${N\to\infty}$.
In contrast, the second-order QPT comes through a discontinuity of the first derivative of the order parameter, which is connected with a discontinuity of the second derivative of the ground-state energy and a polynomially vanishing ground-state energy gap.
In the second-order QPT, the parity of the ground state is spontaneously broken, which means that for ${\lambda>\lambda_{\rm c}(0)}$ and ${\chi=0}$ the ${N\to\infty}$ ground state becomes a degenerate parity doublet.
A more elaborate analysis of the QPTs and the ground-state geometry of Hamiltonian \eqref{H_S} can be found in~\cite{Matu23a,Strel24}.
Finite-$N$ precursors of the QPTs of the first and second order are observed in the evolution of the lowest energy levels in panels (a) and (b) of Fig.\,\ref{fig:fig1}. 
This figure depicts lower parts of energy spectra of our LMG Hamiltonian for two particular paths in the ${\lambda\times\chi}$ plane, which will be used (although with a different value of $N$) in the driving protocols to be introduced in Sec.\,\ref{sec:driving_procedure}.

The ESQPT nonanalyticities, affecting in the ${N\to\infty}$ limit higher energy levels, are also present in the spectrum of Hamiltonian \eqref{H_Lipkin}. 
Their finite-$N$ precursors are also shown in Fig.\,\ref{fig:fig1} (see the parts of the spectra with the colored background).
Associated with the first-order QPT, see Fig.\,\ref{fig:fig1}(a), two chains of avoided level crossings of excited states issue from the ${\lambda\approx\lambda_{\rm c}(\chi)}$ avoided crossing of the ground state to both ${\lambda<\lambda_{\rm c}}$ and ${\lambda>\lambda_{\rm c}}$ directions.
In addition, many other avoided crossings appear in a finite region above the ESQPTs. 
The second-order QPT in Fig.\,\ref{fig:fig1}(b) is accompanied by a single chain of avoided crossings in the direction ${\lambda\gtrsim \lambda_{\rm c}(0)}$.
These structures, which are generic accompaniments of the first- and second-order QPTs~\cite{Cejn08,Cejn21}, play an important role in driving-induced dynamics of the system at nonzero temperature.

\subsection{Environment and qubit-environment coupling}
	
We consider the prototypical environment consisting of a collection of a large number of harmonic oscillators in thermal equilibrium at temperature ${T = 1/\beta}$.
The Hamiltonian (in units of $\varepsilon$) of this heat bath is 
\begin{equation}\label{H_B}
	\hat{H}_{\text{E}} = \sum_{k}\omega_{k}\hat{b}^{\dagger}_{k}\hat{b}_{k},
\end{equation}
where $\hat{b}^{\dagger}_{k}$ and $\hat{b}_{k}$ denote the bosonic creation and annihilation operators and $\omega_k $ are frequencies of individual modes (phonon energies).
To describe the qubit-environment interaction, we follow the Caldeira-Leggett approach~\cite{Cald81,Cald83}, where the qubit system is assumed to be coupled linearly to the position of all the environmental oscillators. 
Therefore, the interaction Hamiltonian (again in units of $\varepsilon$) is expressed as
\begin{equation} \label{H_SB}
    \hat{H}_{\text{I}} = \frac{1}{\sqrt{N}}\ \hat{Q} \otimes
    \underbrace{\sum_{k}g_{k} \big(\hat{b}_{k}^{\dagger} + \hat{b}_{k}\big)}_{\hat{X}},
\end{equation}
where $\hat{X}$ is a weighted position of the oscillators, with~$g_{k}$ denoting non-negative coupling strengths to individual modes, and $\hat{Q}$ is the system-environment coupling operator acting in the space of qubits. 
The operator $\hat{Q}$ is naturally taken as a linear combination of the quasispin components $\hat{J}_{\alpha}$, for instance
\begin{equation}\label{Q}
    \hat{Q}=\sin\theta\,\hat{J}_x+\cos\theta\,\hat{J}_z.
\end{equation}
We use two choices of the angle $\theta$, namely ${\theta=0}$ and ${\theta=\frac{\pi}{2}}$.
While in the former case the environment just modulates energies of the unperturbed qubit states, in the latter case it induces transitions between these states.
Note that with ${\theta=0}$ and ${\chi=0}$ the total Hamiltonian~\eqref{Total_Hamiltonian}  conserves the parity \eqref{parity}, so for ${\chi=0}$ the choice ${\theta=\frac{\pi}{2}}$ captures the generic situation in which the qubit-environment interaction violates non-fundamental symmetries of the qubit system.
The factor in front of the expression \eqref{H_SB} ensures proper scaling of the S-E interaction energy with an increasing number of qubits $N$.

The total Hamiltonian \eqref{Total_Hamiltonian} with individual terms fixed by Eqs.\,\eqref{H_S}, \eqref{H_B} and \eqref{H_SB} suffers from the drawback that for infinite sizes of both the qubit and environment subsystems and for sufficiently large coupling strengths $g_k$ its ground state energy may cease having a lower bound~\cite{Rzaz75}.
In case of a finite $N$, the lower bound exists, but the whole system with increasing $g_k$'s may exhibit a sharp crossover  from a normal phase, characterized by low average occupancies $\ave{\hat{b}_k^{\dagger}\hat{b}_k}$ of all bath modes, to a \lq\lq superradiant\rq\rq\ phase, in which these occupancies get macroscopically large~\cite{Hepp73,Wang73,Emary03}.
This transition is experimentally verified in various laboratory realizations of the Dicke-type Hamiltonians in which the bilinear coupling of the form~\eqref{H_SB} is also present, see, e.g.,~\cite{Baum10,Nata10}.

To avoid the ground-state instability and the transition to the superradiant-like phase, the system Hamiltonian~\eqref{H_S} is often renormalized by adding an extra term, which counterbalances the influence of the environment~\cite{Cald81,Cald83}. 
This leads to a tranformation
\begin{equation}\label{counter}
    \hat{H}_{\rm S}(\bold{\Lambda})\longmapsto 
    \hat{H}'_{\rm S}(\bold{\Lambda},q)=\hat{H}_{\rm S}(\bold{\Lambda})+r\frac{q}{N}\hat{Q}^2,
\end{equation}
where variable $r$ will be explained below and
\begin{equation}\label{strength_q}
    q=\sum_{k}\frac{g_k^2}{\omega_k}
\end{equation}
plays the role of an effective S-E coupling strength.
With the replacement \eqref{counter}, the total Hamiltonian \eqref{Total_Hamiltonian} is guaranteed to have a lower bound of energy~\cite{Kopyl19}.

In this work, the qubit system is always considered finite, so the inclusion of the counterterm is not necessary. 
Nevertheless, as we wish to estimate the effect of renormalization in the driving process, we include the counterterm to some of our simulations.
Whether the renormalization is on or off, respectively, is distinguished by values ${r=1}$ or ${r=0}$ of the auxiliary parameter in Eq.\,\eqref{counter}.
In particular, the option ${r=1}$ is used along with ${\hat{Q}=\hat{J}_x}$, when the coupling term \eqref{H_SB} can be interpreted as the electric dipole interaction of the qubits (atoms carrying the qubit states) with $\omega_k$ modes of a quantized electromagnetic field and the counterterm stands for the $A^2$ term (squared vector potential) of the atom-field Hamiltonian~\cite{Rzaz75,Nata10}.
Hence for ${\theta=\frac{\pi}{2}}$ in Eq.\,\eqref{Q} we use both ${r=0}$ and ${r=1}$ options, while for ${\theta=0}$ we use only ${r=0}$.
With the choice ${r=0}$, we intentionally allow for the situation in which the increasing effective strength~\eqref{strength_q} of the S-E interaction induces a crossover effect in the ground state of the whole system to a superradiant-like phase, which may have a large impact on results of the driving process.

Thermal fluctuations of the environment are described by its self-correlation function
\begin{flalign}\label{correlations} 
    C(t)&=\tr\left[\hat{X}(t)\hat{X}(0)\,\hat{\rho}_{\rm E}^{(\beta)}\right]
    \\
    &= \sum_{k} g_k^2 \biggl[\bigl\{2n_{\rm E}(\omega_k)\!+\!1\bigr\}\cos(\omega_k t) -i\sin(\omega_k t) \biggr] \nonumber\\
     &= \frac{1}{\pi}\int_{0}^{\infty}\!\!\!\! d\omega\,J(\omega) \!\left[\coth\left(\frac{\beta\omega}{2}\right) \!\cos(\omega t) \! - \! i \sin (\omega t) \right],\nonumber
\end{flalign}
where $\hat{\rho}_{\rm E}^{(\beta)}$ is a thermal density operator of the environment and ${\hat{X}(t)=e^{+i\hat{H}_{\text{E}}t}\hat{X}e^{-i\hat{H}_{\text{E}}t}}$ represents time evolution of the  weighted oscillator coordinate from Eq.\,\eqref{H_SB}. 
In the second line we use the Bose distribution $n_{\rm E}(\omega_k) = (e^{\beta\omega_k}-1)^{-1}$ and in the third line we introduce the spectral density
\begin{equation}\label{J}
    J(\omega) = \pi\sum_{k} g_k^2\delta(\omega-\omega_k),
\end{equation} 
which fully determines the correlation function~\cite{Breu02,Weis12}. 
For discrete modes, the spectral density consists of a sequence of $\delta$-peaks, but in the limit of a continuous distribution of bath mode energies (infinite number of bosonic modes) it becomes a smooth function of~$\omega$.
While the real part of the correlation function \eqref{correlations} describes thermal fluctuations and decoherence, the imaginary part captures the effects of dissipation. 
For more detailed explanation we suggest Refs.\,\cite{Breu02,Tani89,Weis12,Schl07}.

It is assumed that the dissipative effect of the bath decays as ${\operatorname{Im}\,C(t) \sim e^{-\gamma t}}$, where ${\gamma^{-1}=\tau_{\text{E}}}$ is the dissipation time scale.
By imposing this condition one finds \eqref{J} in the Drude-Lorentz form~\cite{Tani89,Ma12,Lamb23}:
\begin{equation}\label{J_Drude}
    J(\omega) = 2q\,\frac{\gamma \omega}{\gamma^2 + \omega^2},
\end{equation}
where $\gamma$ represents the width of the bath mode distribution.
With this expression, it is possible to write the discreet sum in \eqref{strength_q} as the integral $q = \pi^{-1}\int_{0}^{\infty}d\omega J(\omega)/\omega$.

The use of this formalism in numerical calculations based on the HEOM method will be further commented in Sec.\,\ref{subsec:Method}.



\section{Driven dynamics}
\label{sec:driving_procedure}

In the driving procedure, the total Hamiltonian \eqref{Total_Hamiltonian} is made time dependent by imposing an external time dependence $\bold{\Lambda}(t)$ of control parameters of the system of qubits.
Thus, ${\hat{H}(t)=\hat{H}_{\text{S}}\bigl(\bold{\Lambda}(t)\bigr)+\hat{H}_{\text{E}}+\hat{H}_{\text{I}}}$.
Below we describe the evolution induced by this Hamiltonian and identify the target states of the qubit system.

\subsection{Initial state and overall evolution}

We assume that the system~S and the environment~E are at the initial time ${t=0}$ prepared with the same inverse temperature $\beta$ in a factorized initial state:
\begin{equation}\label{Initial_State}
    \hat{\rho}(0)=
    \underbrace{\frac{e^{-\beta\hat{H}_{\text{S}}(0) }}{\tr[{e^{-\beta\hat{H}_{\text{S}}(0) }}]}}_{\hat{\rho}_{\text{S}}(0)}
    \otimes
    \underbrace{\frac{e^{-\beta\hat{H}_{\text{E}}}}{\tr[e^{-\beta \hat{H}_{\text{E}}}]}}_{\hat{\rho}_{\text{E}}(0)=\hat{\rho}_{\rm E}^{(\beta)}}.
\end{equation}
Here, $\hat{H}_{\text{S}}(0)$ is the system Hamiltonian at $t=0$, and $\hat{\rho}_{\text{S}}(0)$ and $\hat{\rho}_{\text{E}}(0)$ are thermal density operators of~S and~E, respectively. 
This corresponds to the situation when in times ${t<0}$ both S and E have thermalized separately through interactions with a common thermal reservoir but with no interaction between each other.
We note that the thermal initial state of S in Eq.\,\eqref{Initial_State} is probably a more realistic choice than a pure zero-temperature state ${\hat{\rho}_{\rm S}(0) = \ket{E_0(0)}\bra{E_0(0)}}$. 
Moreover, it will be shown that the nonzero temperature of S has a non-trivial effect on the final fidelity of the driving procedure.

The interaction between S and E is initiated at ${t=0}$, indicating the start of the driving procedure on S, see Sec.\,\ref{dripro}.
One can assume that the environment E is connected with the instrumentation needed to perform the drive.
Since then, the S-E interaction completely dominates over the interaction with the other reservoir so that the latter can be neglected.
The evolved state $\hat{\rho}(t)$ of the composite S-E system for ${t>0}$ satisfies the quantum Liouville--von Neumann equation
\begin{equation}\label{liov}
\frac{d}{dt}\hat{\rho}(t)=-i\left[\hat{H}(t),\hat{\rho}(t)\right].
\end{equation} 
It is very probable that because of the interaction between S and E, the total density operator $\hat{\rho}(t)$ is no more factorized.
The density operator of the qubit system is then extracted by the common procedure involving the partial trace over the Hilbert space of the environment:
\begin{equation}\label{partial}
   \hat{\rho}_{\text{S}}(t)={\rm tr}_{\rm E}\,\hat{\rho}(t).
\end{equation} 
We outline the method how this evolution is determined in Sec.\,\ref{subsec:Method}.
Before, we focus on the externally driven variation of the system Hamiltonian $\hat{H}_{\rm S}\bigl(\bold{\Lambda}(t)\bigr)$ and on the definition of the state of the qubit system which is the target of our driving procedure.

\subsection{Driving protocols}
\label{dripro}

The time-dependent system Hamiltonian $\hat{H}_{\rm S}\bigl(\bold{\Lambda}(t)\bigr)$ follows from a predefined time dependence of control parameters $\bold{\Lambda}$ of Hamiltonian \eqref{H_S}.
The function $\bold{\Lambda}(t)$ is determined by the chosen path in the parameter space and by the speed with which the system is driven along this path.
In our case, the path will always be a straight line connecting the initial and final parameter points $\bold{\Lambda}_{\rm I}$ and $\bold{\Lambda}_{\rm F}$, respectively. 
This line is parametrized as
\begin{eqnarray} \label{linear_path}
    \bold{\Lambda}(t) = \bold{\Lambda}_{\rm I} + 
    s(t) \left( \bold{\Lambda}_{\rm F} - \bold{\Lambda}_{\rm I} \right),
\end{eqnarray}
where $s(t)$ is a variable satisfying ${s(0)=0}$ and ${s(t_{\rm F})=1}$, with $t_{\rm F}$ denoting the final time, i.e., the total time of the driving.
This ensures that ${\bold{\Lambda}(0)=\bold{\Lambda}_{\rm I}}$ and ${\bold{\Lambda}(t_{\rm F})=\bold{\Lambda}_{\rm F}}$. 
Moreover, we assume a continuous and monotonous variation of $s(t)$ between its initial and final values 0 and 1, and consider the following two options for its time dependence.

\begin{figure}[tp]
    \includegraphics[width=1.\linewidth]{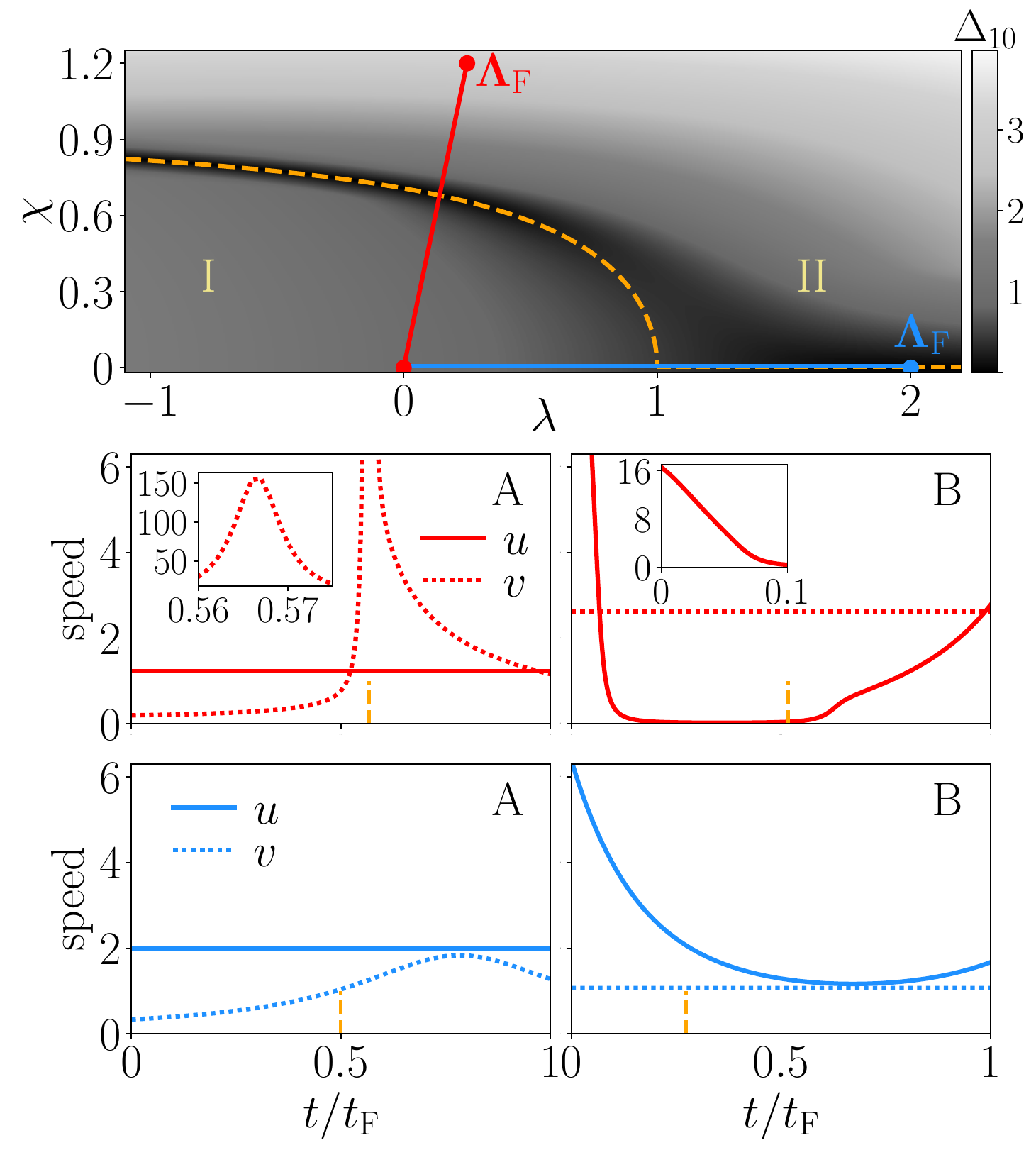}
    \caption{Top panel: Relevant part of the parameter space of the qubit Hamiltonian $\hat{H}_{\rm S}(\bold{\Lambda})$. The dashed curve represents the ${N \rightarrow \infty}$ QPT separatrix between the ground-state phases I and II. The tilted (solid red) line segment represents the driving path across the first-order QPT. The horizontal (solid blue) line segment represents the driving path across the second-order QPT. The gray background encodes the energy gap $\Delta_{10}(\bold{\Lambda})$ for ${N = 10}$.   
    Lower panels: The planar speed~$u$ (solid line) and geometric speed~$v$ (dotted line) for driving protocols A and B along both paths for ${N = 10}$ and ${t_{\rm F} = 1}$. The upper and lower rows (with red and blue curves) correspond to the paths across the first- and second-order QPTs, respectively (overflow parts of the dependencies are displayed in the insets). The vertical dashed line segments mark the crossing of the ${N \rightarrow \infty}$ QPT separatrix from the upper panel.}
    \label{fig:Parameter_space}
\end{figure}

Protocol A: The first option is the simplest linear dependence $s(t)=t/t_{\rm F}$. 
This ensures that the speed on the parameter plane
\begin{equation}\label{speed_u}
   u(t)=\sqrt{\sum_{\mu,\nu}\delta_{\mu\nu}\frac{d\Lambda^{\mu}}{dt}\frac{d\Lambda^{\nu}}{dt}},
\end{equation}
hereafter referred to as the \lq\lq planar speed\rq\rq, is kept constant, equal to ${u=\left| \bold{\Lambda}_{\rm F} - \bold{\Lambda}_{\rm I} \right|/t_{\rm F}}$. 

Protocol B: The second option is a more complicated time dependence of parameter $s(t)$, that keeps constant the so-called \lq\lq geometric speed\rq\rq 
\begin{equation}\label{speed_v}
v(t)=\sqrt{\sum_{\mu,\nu}g_{\mu\nu}\bigl(\bold{\Lambda}(t)\bigr)\frac{d\Lambda^{\mu}}{dt}\frac{d\Lambda^{\nu}}{dt}}.
\end{equation}
Here, $g_{\mu\nu}(\bold{\Lambda})$ is the metric tensor defined on the manifold of ground states $\ket{E_0(\bold{\Lambda})}$ of Hamiltonian $\hat{H}_{\rm S}(\bold{\Lambda})$ 
via the distance element ${dl^2=1-\left|\scal{E_0(\bold{\Lambda})}{E_0(\bold{\Lambda}\!+\!d \bold{\Lambda})}\right|^2}$~\cite{Prov80}.
Assuming a non-degenerate spectrum for finite $N$, we can write 
\begin{equation}\label{metric}
    g_{\mu\nu} = {\rm Re}\sum_{n>0} \frac{\matr{E_0}{\frac{\partial}{\partial\Lambda^{\mu}}\hat{H}_{\text{S}}}{E_n} \matr{E_n}{\frac{\partial}{\partial\Lambda^{\nu}}\hat{H}_{\text{S}}}{E_0}}{(E_n-E_0)^2},
\end{equation}
where for brevity we skip all dependencies on~$\bold{\Lambda}$.

The metric~\eqref{metric} and the geometric speed~\eqref{speed_v} reflect variations of the ground-state wave function with control parameters $\bold{\Lambda}$.
In protocol B, the geometric speed is kept constant, equal to $v=\int_{\bold{\Lambda}_{\rm I}}^{\bold{\Lambda}_{\rm F}}\sqrt{g_{\mu\nu}d\Lambda^{\mu}d\Lambda^{\mu}}/t_{\rm F}$, while the planar speed \eqref{speed_u} varies in time, taking smaller values in the parameter domains where~$g_{\mu\nu}(\bold{\Lambda})$ is large, i.e., particularly in regions with a small energy gap $\Delta_{10}(\bold{\Lambda})$.
For drivings of isolated systems initiated in the ground state $\ket{E_0(\bold{\Lambda}_{\rm I})}$, the $v$=const. strategy improves the final fidelity for the target state $\ket{E_0(\bold{\Lambda}_{\rm F})}$~\cite{Matu23a,Matu23b} (see also~\cite{Rola02} for a closely related method). 
We will show that this advantage lasts even at some nonzero (sufficiently low) temperatures.
In principle, one can go beyond protocol~B by using a curved path that minimizes the geometric length from $\bold{\Lambda}_{\rm I}$ to $\bold{\Lambda}_{\rm F}$, but this approach was shown to yield unsatisfactory results~\cite{Matu23a} and we do not apply it here. 

The initial and final points in the $\bold{\Lambda}=(\lambda,\chi)$ plane define two driving paths:
\begin{eqnarray}
  \bold{\Lambda}_{\rm I} \!\!&=&\!\!\! (0,0),
  \label{ini}\\
  \bold{\Lambda}_{\rm F} \!\!\!&=&\!\!\! \left\{ \begin{array}{ll} \!\! (0.25,1.2) &\!\text{path across first-order QPT,} \\ \!\! (2,0) & \!\text{path across second-order QPT.} \end{array} \right.
  \quad
  \label{fin}
\end{eqnarray}
This means that the driving always starts at $\bold{\Lambda}_{\rm I}$ corresponding to the factorized ground state of the system of independent qubits.
In contrast, both final points $\bold{\Lambda}_{\rm F}$ in Eq.\,\eqref{fin} are associated with an entangled ground state of the system of strongly interacting qubits. 
For the first choice of the final point, the driving trajectory crosses the first-order QPT, the parity \eqref{parity} being broken.
For the second choice the trajectory crosses the second-order QPT along the parity-conserving line with ${\chi=0}$.
The crossing of the respective critical point (its ${N\to\infty}$ realization) corresponds to ${s\approx 0.57}$ for the first-order QPT path and ${s=0.5}$ for the second-order QPT path.

In Fig.\,\ref{fig:Parameter_space} we plot the two driving paths in the plane $(\lambda,\chi)$ and the speeds \eqref{speed_u} and \eqref{speed_v} along both paths for driving protocols A and B for the system with ${N=10}$ qubits.
We note the reduction of the planar speed $u(t)$ in both first- and second-order QPT regions for the $v$=const. driving procedure B.
This slowdown is correlated with a reduced ground-state energy gap $\Delta_{10}$, the minimal gap for ${N=10}$ being however shifted with respect to the ${N\to\infty}$ QPT separatrix.

\subsection{Target states and fidelity}

The target state for any of the above-described driving procedures is the ground state of the qubit system at the final parameter point ${\bold{\Lambda}=\bold{\Lambda}_{\rm F}}$.
For the path across the first-order QPT with ${\bold{\Lambda}_{\rm F}=(0.25,1.2)}$, the final ground state is given by a single eigenvector $\ket{E_0(\bold{\Lambda}_{\rm F})}$, which is energetically well separated from the other eigenvectors. 
So, if $\hat{\rho}_{\text{S}}(t_{\rm F})$ is the density operator of the qubit system at the end of the driving procedure, see Eq.\,\eqref{partial}, then the fidelity of the target-state preparation is
\begin{equation}\label{fidel1} 
    \mathcal{F}_1=\matr{E_0(\bold{\Lambda}_{\rm F})}{\hat{\rho}_{\text{S}}(t_{\rm F})}{E_0(\bold{\Lambda}_{\rm F})}.
\end{equation}
This formula expresses the probability of finding the target ground state $\ket{E_0(\bold{\Lambda}_{\rm F})}$ in the statistical ensemble associated with the final density operator $\hat{\rho}_{\text{S}}(t_{\rm F})$.

For the ${\chi=0}$ path across the second-order QPT with ${\bold{\Lambda}_{\rm F}=(2,0)}$, the final ground state in the ${N\to\infty}$ limit is a degenerate doublet of positive- and negative-parity states.
Although the degeneracy of the positive-parity eigenstate $\ket{E_0(\bold{\Lambda}_{\rm F})}$ and the negative-parity eigenstate $\ket{E_1(\bold{\Lambda}_{\rm F})}$ is not exact for finite $N$, we calculate the final fidelity by summing the overlaps with both these states:
\begin{eqnarray}\label{fidel2}
    &\mathcal{F}_2={\rm tr} \left[\hat{P}_{{\rm gs}}(\bold{\Lambda}_{\rm F})\,\hat{\rho}_{\text{S}}(t_{\rm F})\right]=
\\  
    &\matr{E_0(\bold{\Lambda}_{\rm F})}{\hat{\rho}_{\text{S}}(t_{\rm F})}{E_0(\bold{\Lambda}_{\rm F})}
    +\matr{E_1(\bold{\Lambda}_{\rm F})}{\hat{\rho}_{\text{S}}(t_{\rm F})}{E_1(\bold{\Lambda}_{\rm F})},
\nonumber
\end{eqnarray}
where $\hat{P}_{{\rm gs}}(\bold{\Lambda}_{\rm F})$ is the projector to the subspace spanned by vectors $\ket{E_0(\bold{\Lambda}_{\rm F})}$ and $\ket{E_1(\bold{\Lambda}_{\rm F})}$.
This formula expresses the probability that any state randomly drawn from the statistical ensemble associated with $\hat{\rho}_{\text{S}}(t_{\rm F})$ lies in the quasidegenerate ground-state subspace at ${\bold{\Lambda}=\bold{\Lambda}_{\rm F}}$.
	
\subsection{Notes on the HEOM calculations}
\label{subsec:Method}

The evolution of the qubit density operator $\hat{\rho}_{\text{S}}(t)$ from Eq.\,\eqref{partial} is determined by the HEOM method~\cite{Tani89,Tani06,Fruch16,Tani20,Lamb23}, which is capable to yield exact (nonperturbative) numerical description of the system dynamics.
In particular, it goes beyond the approximations involved in the common Lindblad formalism, which strictly relies on the assumption that environment-induced processes on the system are Markovian. 
The method originates from the influence functional formalism of Feynman and Vernon~\cite{Feyn63},
which arises from the integration over all degrees of freedom of the environment and determines the system evolution $\hat{\rho}_{\text{S}}(t)$ on the basis of the system self-Hamiltonian $\hat{H}_{\rm S}$, the coupling operator $\hat{Q}$ from Eq.\,\eqref{H_SB} and the environment self-correlation function $C(t)$ from Eq.\,\eqref{correlations}.

We do not immerse here deeper into the description of the HEOM method but just present its very rough outline and introduce the parameters involved.  
The evolution $\hat{\rho}_{\text{S}}(t)$ is described by a set of coupled first-order differential equations~\cite{Tani89}, which can be efficiently solved in terms of some auxiliary density matrices.
These matrices do not represent real physical states but serve just as a numerical tool to account for the S-E correlations depending on the coupling strength $q$, temperature $T$ and dissipation time scale $\tau_{\rm E}$.
When $q$ is small and $\tau_{\rm E}$ much lower than the internal dynamical time scale of the system, the dynamics is approximately Markovian and only a few auxiliary density matrices are needed. 
In contrast, for large $q$ and $\tau_{\rm E}$ the dynamics is non-Markovian and a large number of matrices is required to account for the long-lasting S-E correlations. 
The auxiliary matrices are grouped under a certain hierarchy of layers enumerated by ${0,1,\dots,L}$, which are connected with the inclusion of time correlations of an increasing order.
The depth~$L$ of this hierarchy must be chosen large enough to ensure the convergence.
Details can be found, e.g., in Ref.\,\cite{Tani06}.

In order to solve the set of HEOM equations, the self-correlation function \eqref{correlations} is expanded as a series of exponentials.
This can be done analytically, in general, for a spectral density $J(\omega)$ that admits a finite number of poles in the complex plane and that goes to zero as $\omega\rightarrow\infty$. 
For the Drude-Lorentz distribution \eqref{J_Drude} the series takes the form
\begin{equation}\label{serc}
C(t) = c_0\, e^{-\gamma t}+\sum_{k=1}^{\infty} c_k\, e^{-2\pi k T t}
\end{equation}
with coefficients ${c_0\in{\mathbb C}}$ and ${c_1,c_2,\ldots\in{\mathbb R}}$ depending on parameters $q$ and $\gamma$, and on the temperature $T$. 
The explicit expressions are given, e.g., in Ref.\,\cite{Lamb23}.
In numerical calculations the sum in Eq.\,\eqref{serc} can be truncated at its $M$th term. 
For low temperatures, the convergence is slow and $M$ must be relatively large to reach an acceptable precision of calculations. 
On the other hand, for high temperatures even very small $M$ may lead to reasonable results. 
In any case, to compensate this truncation, a Lindbladian term that speeds up the convergence (so-called terminator) is included~\cite{Ishi05}.


During the last decades, several open-source libraries have been developed to implement general calculations based on the HEOM formalism. 
Here we use the version found in the Python library QuTiP-BoFiN, see Ref.\,\cite{Lamb23}. 


\section{RESULTS}
\label{sec:results_discussion}

In our calculations we use the value ${\gamma = 10}$ for the width of the Drude-Lorentz distribution \eqref{J_Drude} in units of~$\varepsilon$.
This means that the typical time scale of the environment is smaller (10 times for ${\lambda=\chi=0}$) than that of the qubit system.
We consider three values (again in units of $\varepsilon$) of the effective system-environment coupling strengths: ${q = 0}$ (no coupling), ${q = 0.1}$ (medium coupling) and ${q=1}$ (strong coupling).
The above nomenclature concerning the coupling strength should not be taken literally as the effect of the coupling is expected to depend on the driving path.
In particular, for the path across the first-order QPT both ${q\neq 0}$ values represent relatively strong coupling regimes (the case ${q=1}$ may even be called an ultrastrong coupling), while for the path across the second-order QPT the case ${q=0.1}$ corresponds to a weak coupling regime. 
In most calculations, we set the size of the qubit system to a relatively low value, namely ${N = 10}$.
In that case, the width $\gamma$ of the Drude-Lorentz distribution covers the whole spectrum of the qubit system with ${\lambda=\chi=0}$.

At low temperatures, values of the truncation parameter up to ${M=18}$ are needed to ensure the convergence of the HEOM calculations, while at high temperatures, the value ${M=5}$ is found to be sufficient.
For all temperatures, the number of layers is set to ${L=3}$, the contributions of the ${L \geq 4}$ layers being shown to yield only negligible corrections.
This holds even at the low temperatures, when a short period of ${{\rm Re}\,C(t)<0}$ induces a strong non-Markovian behavior.

At very high temperatures, ${T \gtrsim 25}$, all states of the qubit system become almost equally populated. 
Thus, independently of the coupling strength $q$ and the driving time $t_{\rm F}$, the fidelity for drivings across the first- and second-order QPT approaches the values 
\begin{equation}\label{limF}
    \mathcal{F}_1\xrightarrow[T\to\infty]{}\frac{1}{N+1},\quad \mathcal{F}_2\xrightarrow[T\to\infty]{}\frac{2}{N+1},
\end{equation}
where $(N+1)$ is the dimension of the qubit Hilbert space. 
Therefore, we consider the fidelity for different driving times $t_{\rm F}$ as a function of temperature ${T \leq 25\approx 10^{1.4}}$.  

Results for the first- and second-order QPT driving paths are described separately. 
A short comparison of the first-order QPT results obtained by the HEOM calculations with those based on the Lindblad formalism is presented afterwards.

\subsection{First-order QPT}
\label{se:first}

\begin{figure*}[tp]
    \includegraphics[width=0.85\linewidth]{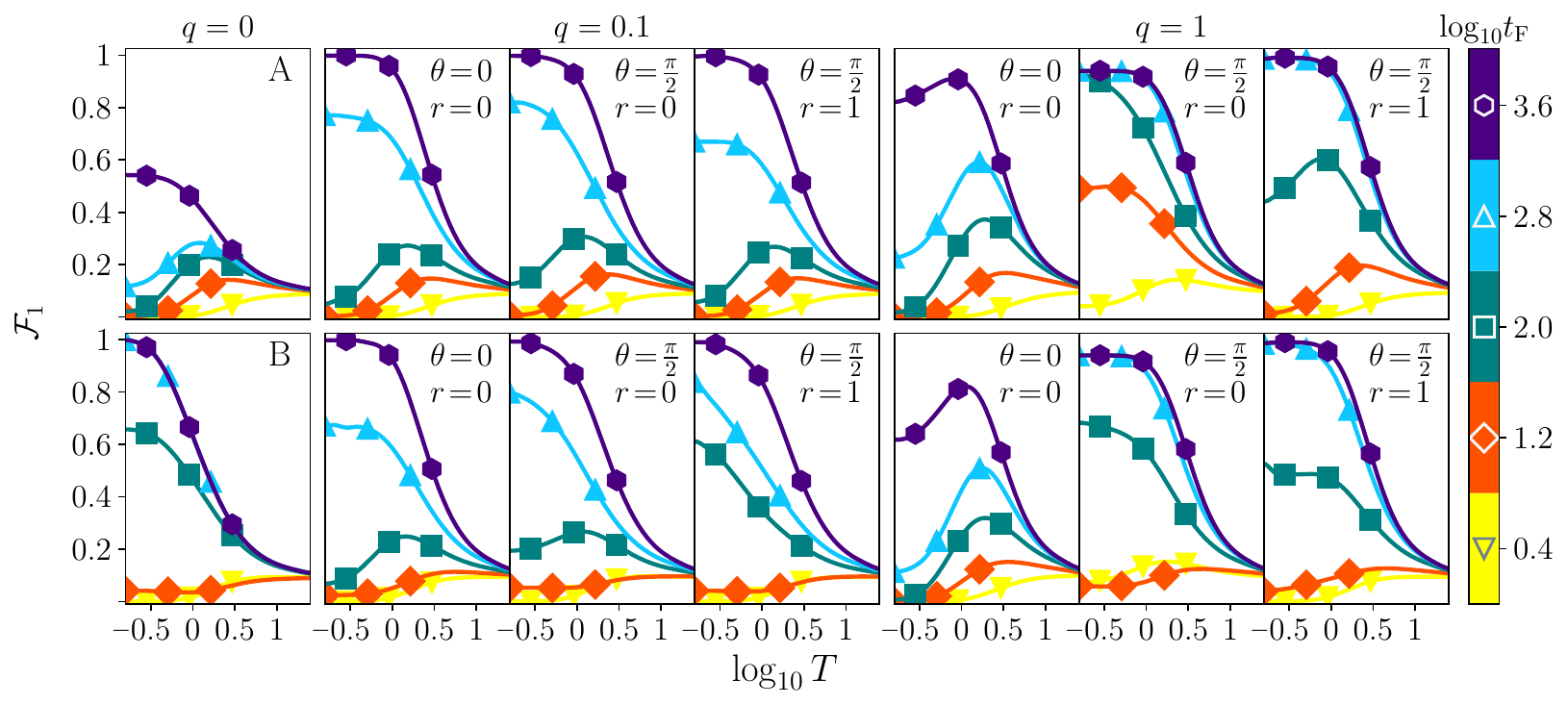}
    \caption{Final fidelity of the ${N=10}$ qubit system for drivings across the first-order QPT precursor. Results for driving protocols A and B are shown in the top and bottom rows, respectively. The effective coupling $q$ from \eqref{strength_q}, the angle $\theta$ in \eqref{Q} and the parameter ${r}$ in \eqref{counter} are indicated above or within each subplot. Each color and marker type represents a distinct value of the driving time $t_{\rm F}$, as indicated in the bar on the right (the markers are used just to label the curves and do not correspond to all calculated points). The range of temperature is $\log_{10}T\in[-0.8,1.4]$.
    }
    \label{fig:First_order_fidelity}
\end{figure*}

Results of calculations of the target state fidelity $\mathcal{F}_1$ from Eq.\,\eqref{fidel1} for drivings across the first-order QPT are summarized in Fig.\,\ref{fig:First_order_fidelity}.
The two rows of the figure correspond to driving protocols A and B (see Sec.\,\ref{dripro}), columns in each row present results for the three values of the effective system-environment coupling strength $q$ given above.
For both $q\neq 0$ values (the medium and strong coupling cases) we distinguish the two choices of the coupling operator $\hat{Q}$ from Eq.\,\eqref{Q} by the angle $\theta$. The inclusion of the counterterm in \eqref{counter} is indicated by ${r=1}$ ($r=0$ implies no addition of it).
Individual curves in each plot correspond to distinct values of the driving time $t_{\rm F}$. 
These are distinguished by colors and markers that will be used also in some forthcoming figures.

Figure~\ref{fig:First_order_fidelity} contains rather complex information about the dependence of the fidelity on numerous variables and conditions.
We start to disentangle this information by picking up some basic trends:
\begin{itemize}
\item[(a)] Quite expectedly, slow drivings (large times $t_{\rm F}$) lead in general to higher fidelity than fast drivings (small $t_{\rm F}$). The difference is substantial for low temperatures $T$ and weakens for higher~$T$. 
\item[(b)] In absence of coupling to the environment (${q=0}$), protocol B leads to higher fidelity than A.
\item[(c)] In the case of protocol A and sufficiently long times $t_{\rm F}$, the onset of interaction with the environment (${q\neq 0}$) leads to a considerable increase of fidelity. The effect is present for both medium and strong couplings, being more pronounced for ${\theta=\frac{\pi}{2}}$ and ${r=0}$. For protocol B the effect is absent, so the above-mentioned generally better performance of B does not hold for ${q>0}$.
\item[(d)] Whereas for low and medium temperature $T$ the fidelity depends (more or less sensitively) on all the above-mentioned variables and conditions, for high temperatures we observe convergence of all curves to the uniform limit \eqref{limF}. 
\item[(e)] We observe nonmonotonous dependencies of some of the $\mathcal{F}_1$ curves on $T$ for both driving types.
\item[(f)] Effects caused by the inclusion of the counterterm (${r=1}$) in Eq.\,\eqref{counter} depend on the coupling strength: For ${q=1}$, we observe a systematic increase of fidelity (compared to the corresponding ${r=0}$ cases) for long times~$t_{\rm F}$ and a decrease for shorter times. This holds for both driving types. For ${q=0.1}$ the effect of the counterterm depends nonsystematically on the driving type and time. 

\end{itemize}

If one seeks a way how to increase the fidelity of the target ground-state preparation, the above observations can be translated to the following practical instructions:

First, based on item (a): {\it Cool down the system and set a large driving time.}
This strategy is just a straightforward extension of the common notion of adiabaticity from isolated to open systems.

Second, based on item (b): {\it If the system is nearly isolated, use the driving protocol {\rm B} instead of {\rm A}.}
Indeed, for ${q=0}$ the two slowest drivings with protocol B reach almost the perfect fidelity $\mathcal{F}_1\approx 1$, while with protocol~A the fidelity barely approaches to 0.54 and 0.12 for the same driving times.
This instruction, which is consistent with the results of Ref.\,\cite{Matu23a}, follows from the reduction of the planar speed $u(t)$ in the QPT region with a small ground-state energy gap, see Fig.\,\ref{fig:Parameter_space}, which suppresses transitions to higher-energy levels.

\begin{figure*}[tp]
   \includegraphics[width=0.85\linewidth]{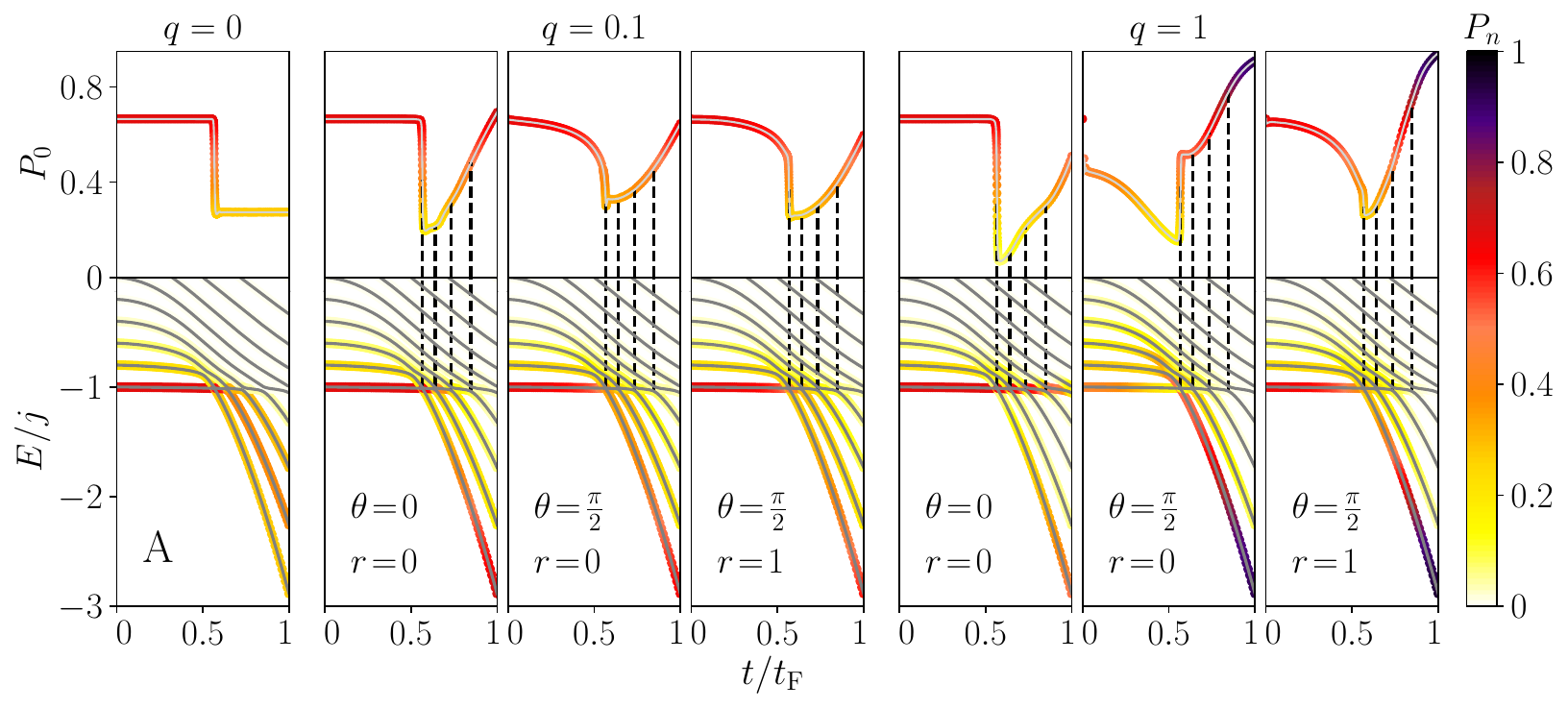}
    \caption{Evolving occupation probabilities $P_n(t)$ of instantaneous energy eigenstates ($n=0,1,\ldots$) of the ${N=10}$ qubit system for driving protocol A along the path crossing the first-order QPT with ${t_{\rm F}=10^{2.8}}$ and ${T=10^{-0.04}}$. The choices of $q$, $\theta$ and $r$ are indicated. The bottom row depicts the whole low-lying part of the spectrum, with $P_n(t)$ encoded in the color of the respective level (see the color bar). The top row shows the occupation probability of the ground state. The vertical dashed lines mark the avoided crossings of the ground and excited states, which are correlated with changes of the occupation probabilities.}
		\label{fig:Instantaneous_occupation}
\end{figure*}

The third instruction is an alternative to the second one and issues from item (c): {\it Stick with driving protocol~{\rm A} but let the system interact with the environment.}
We see that the slowest driving of type A with the S-E coupling strength ${q=0.1}$ reaches the fidelity $\mathcal{F}_1\approx 0.99$ at low temperature, comparable to driving B with ${q=0}$, but in this case the effect survives even to medium temperatures.
The increase of fidelity is present even for the strong coupling ${q=1}$ , although in a less distinct form and also depending on the coupling operator~$\hat{Q}$. 
For the strong coupling, the counterterm in Eq.\,\eqref{counter} decreases the enhancement for fast drivings but increases it for slower drivings.
The increase of the target ground-state population in these cases is due to the dissipation and ongoing thermalization of the system during the drive.
The same mechanism apparently does not work for the driving protocol~B.
In this case, the fidelity with ${q=0}$ is already high and the onset of nonzero coupling washes out the advantage of the ${v={\rm const.}}$ strategy because the S-E interaction populates excited states of the system (which makes the ground-state metric less relevant than in the ${q=0}$ case).

To unfold the mechanism of the environment-induced increase of fidelity, we show in Fig.\,\ref{fig:Instantaneous_occupation} an example of instantaneous populations of the low-lying states in the spectrum of the qubit system during the driving procedure A for some particular temperature and driving time.
The occupation probability of the $n$th state is given by 
\begin{equation}\label{occupy}
 P_n(t)=\big\langle E_n(\bold{\Lambda}(t))\big|\hat{\rho}_{\text{S}}(t)\big|E_n\big(\bold{\Lambda}(t)\big)\big\rangle.   
\end{equation}
So $P_0(t)$ is the instantaneous occupation of the ground state, $P_1(t)$ is the instantaneous occupation of the first excited state, and so on.
For ${q=0}$ only transitions of the Landau-Zener type modify instantaneous populations of individual states, which are initially set to the thermal values ${P_n(0)\propto e^{-\beta E_n(\bold{\Lambda}_{\rm I})}}$.
We observe that in this case the ground-state occupation probability $P_0(t)$ sharply drops at the QPT-related ground-state avoided level crossing.
However, for ${q>0}$, the dissipation and thermalization processes induced by interactions with the environment gradually push the ground-state occupation probability to higher values after the QPT crossing, in some cases going even above the initial population $P_0(0)$.
This holds for both ${r=0}$ and 1 cases.

The last and maybe most surprising strategy for increasing the target ground-state fidelity comes from the observation mentioned in the above item (e): {\it Increase the temperature to the value for which $\mathcal{F}_1$ is maximal.}
Indeed, many curves in Fig.\,\ref{fig:First_order_fidelity}, particularly those corresponding to smaller driving times, show a clear maximum at a certain optimal temperature ${T_{\rm opt}>0}$.
The increase of fidelity reached in this way is not huge (and sometimes no increase is even present), but in some cases it is not negligible, like for driving A with ${t_{\rm F}=10^{2.8}}$ in the strong coupling case with ${\theta=0}$, when one goes from ${\mathcal{F}_1\approx 0.2}$ at low $T$ to ${\mathcal{F}_1\approx 0.6}$ at ${T=T_{\rm opt}}$.
While for driving protocol~A this effect is present for all values of~$q$, for protocol~B it appears only for ${q>0}$.

\begin{figure}[tp]
    \includegraphics[width=\linewidth]{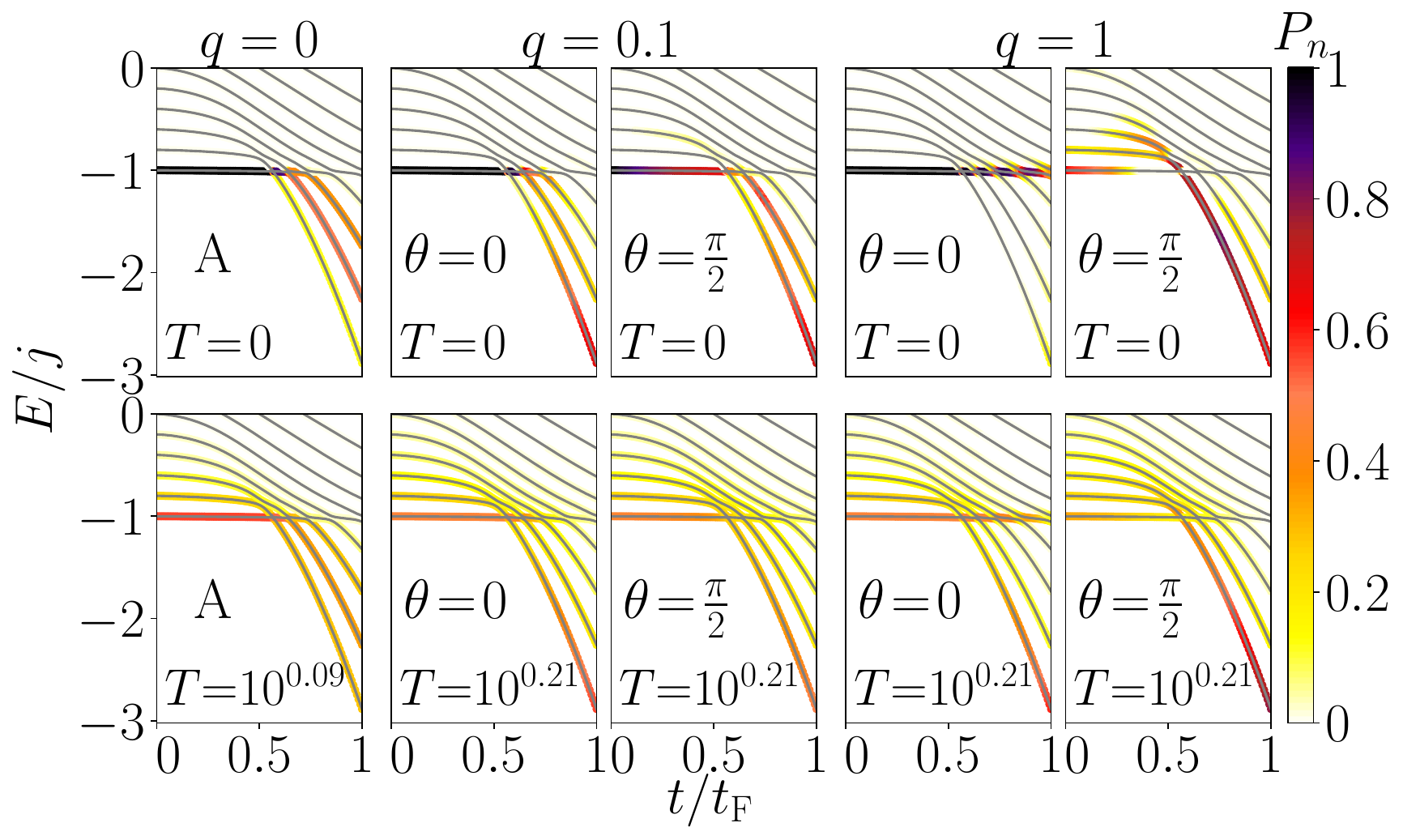}
    \caption{Occupation probabilities $P_n(t)$ of instantaneous low-energy eigenstates of the ${N=10}$ qubit system for driving protocol A along the path across the first-order QPT with ${t_{\rm F}=10^{2.8}}$ at temperatures $T$ specified in each plot. The values of $q$ and $\theta$ are indicated, ${r=0}$.  All plots in the top row correspond to zero temperature. In the bottom row, the temperature is set to the respective optimal value $T_{\rm opt}$ for the ${q = 0}$ and ${q = 1}$, ${\theta = 0}$ plots. In the other cases, for which the dependence of $\mathcal{F}_1$ on $T$ in Fig.\,\ref{fig:First_order_fidelity} is monotonous, $T$ is set to the same value as in the latter case.}
    \label{fig:Zero_T_occupation}
\end{figure}

The nature of this effect is studied in Fig.\,\ref{fig:Zero_T_occupation}.
It compares evolving occupation probabilities \eqref{occupy} of individual energy eigenstates of the qubit system for protocol~A with a particular final time $t_{\rm F}$, the driving being performed at different temperatures~$T$ (this time we only show results without the counterterm). 
We again (as in Fig.\,\ref{fig:Instantaneous_occupation}) observe changes of level populations induced by transitions of the Landau-Zener type and those caused by interactions with the environment.
In the top row of plots, which all correspond to ${T=0}$, the depopulation of the ground state in the QPT region via the Landau-Zener mechanism plays an important role for the final fidelity.
An interesting exception is the case with ${q=1}$ and ${\theta = \frac{\pi}{2}}$, where energy transfers from the environment excite the system even far before the QPT.
The bottom row of plots corresponds to the same driving at higher temperatures.
We notice that in this case, the thermal population of excited states can sometimes increase the population of the ground state right after the QPT.
This happens via the same Landau-Zener mechanism, which now has a partly positive effect.
This effect is even enhanced by transfers of populations to lower excited states when passing the ESQPT-related sequence of avoided-crossings before the QPT.
In particular, for ${q=0}$ and for ${q=1}$, ${\theta=0}$ we set $T$ to its respective optimal value $T_{\rm opt}$ and observe an improvement of the final fidelity.
However, because of complex interplay between all effects involved in the evolution of individual populations, the increase of temperature does not always lead to an advantage, as seen in Fig.\,\ref{fig:First_order_fidelity}. 
Indeed, for the other choices of $q$ and $\hat{\theta}$ in Fig.\,\ref{fig:Zero_T_occupation}, for which $\mathcal{F}_1$ in Fig.\,\ref{fig:First_order_fidelity} monotonously decrease with $T$, we detect a lowering of the ground-state population at the end of the driving procedure in comparison with the $T=0$ case.

\begin{figure}[t]
    \includegraphics[width=\linewidth]{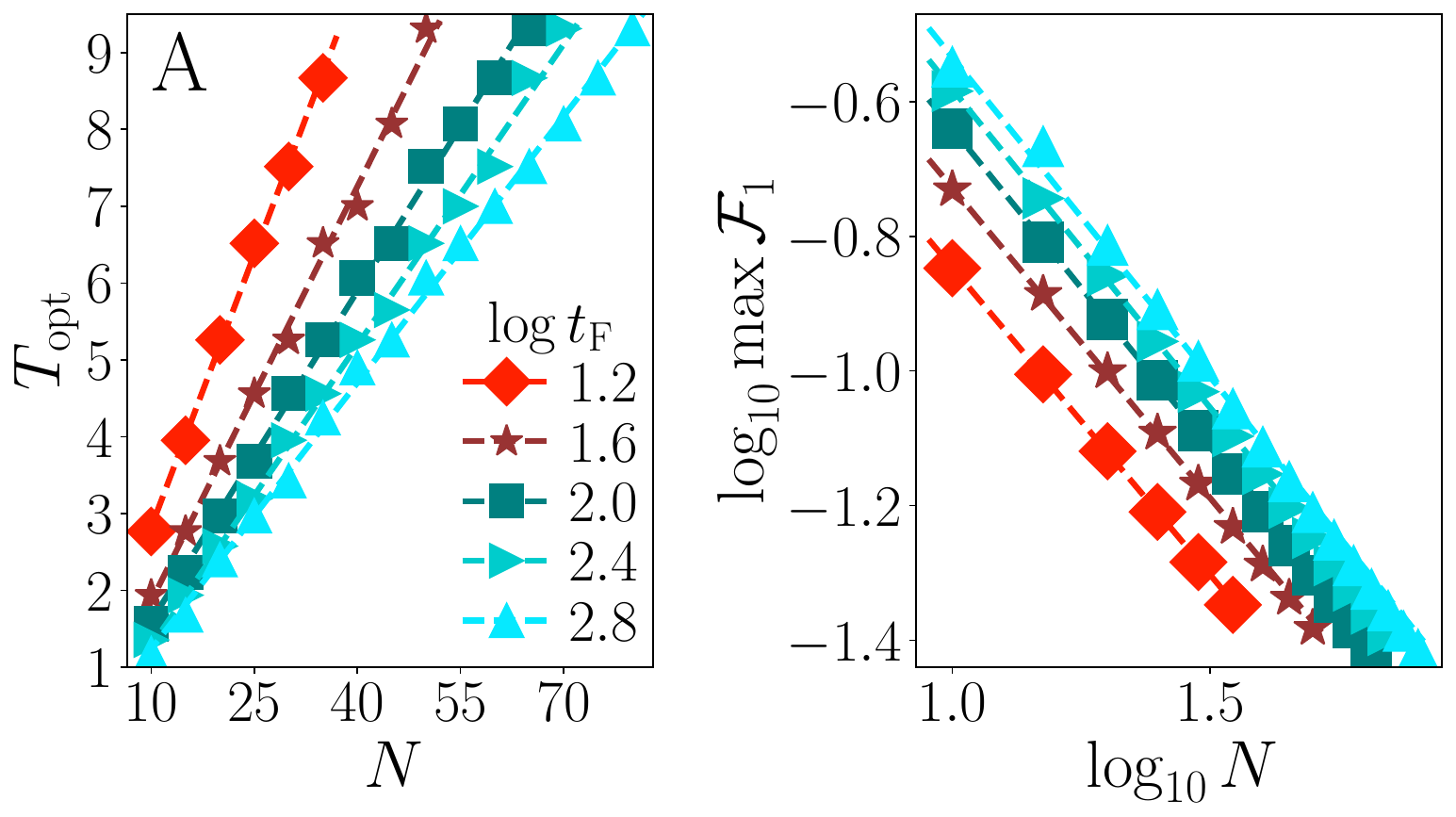}
    \caption{Optimal temperature $T_{\rm opt}$ (left panel) and the fidelity ${\rm max}\,\mathcal{F}_1$ obtained for this temperature (right panel) as functions of the size parameter $N$ for driving protocol A across the first-order QPT with ${q=0}$. The markers represent numerical data for different driving times $t_{\rm F}$ (we use the same marker colours and types as in Fig.\,\ref{fig:First_order_fidelity}), while the dashed lines are linear fits.}
    \label{fig:T_opt_andmore}
\end{figure}

Motivated by the experimental work in Ref.\,\cite{Alba17}, which analyzed how the maximal temperature needed to reach fidelity at or above a certain limiting value scales with the size of the system, we further investigate the dependence of the above-discussed temperature-driven effect on the number $N$ of qubits.
In Fig.\,\ref{fig:T_opt_andmore} we show the $N$-dependence of the optimal temperature~$T_{\rm opt}$ and of the locally maximal value of fidelity ${\rm max}\,\mathcal{F}_1$ at this temperature.
We use protocol A with different driving times $t_{\rm F}$ and set ${q=0}$ for simplicity.
We see that the optimal temperature grows approximately linearly with~$N$, hence ${T_{\rm opt}=aN}$ with the coefficient $a$ depending on $t_{\rm F}$. 
On the other hand, the local maximum of fidelity at $T_{\rm opt}$ drops algebraically, ${{\rm max}\,\mathcal{F}_1=bN^{-\kappa}}$ with $b$ depending on~$t_{\rm F}$ and ${{\kappa\approx 1}}$.
It is obvious that this finding strongly reduces the applicability of the last fidelity-increase strategy in systems with a larger number of qubits.

In any case, nonmonotonous dependencies of fidelity on temperature have already been detected experimentally in Ref.~\cite{Dick13}.
Although the mechanism of the thermally assisted quantum annealing discussed in that study is basically the same as the mechanism of the present effect, some aspects of both systems are different.
In particular, the setup of Ref.~\cite{Dick13} involved a single level crossing including only the ground and first excited states, while the higher energy eigenstates were just spectators.
In that situation, it was possible to derive an analytic estimate of the optimal temperature $T_{\rm opt}$.
In our case, however, that estimate does not work because of the more complicated ESQPT-related structure of avoided crossings.

Let us finally note that our present results (see particularly the ${T=0}$ row of Fig.\,\ref{fig:Zero_T_occupation}) can be compared with the previous studies~\cite{Wubs06,Yama17} of driving of a dissipative single-qubit (two-level) model through a single avoided level crossing at zero temperature.
The system-environment coupling operators employed in these studies were identical with our choice, namely ${\hat{Q}\propto\hat{\sigma}_z}$ and ${\hat{Q}\propto\hat{\sigma}_x}$.
Although some of our findings are compatible with these older results (in particular, the increase of the ground-state population with the onset of dissipation~\cite{Yama17}), our present study demonstrates that the ${N>1}$ qubit system yields more complex dependencies than the simplest case of ${N=1}$.

\subsection{Second-order QPT}
\label{se:second}

\begin{figure*}[tp]
    \includegraphics[width=0.85\linewidth]{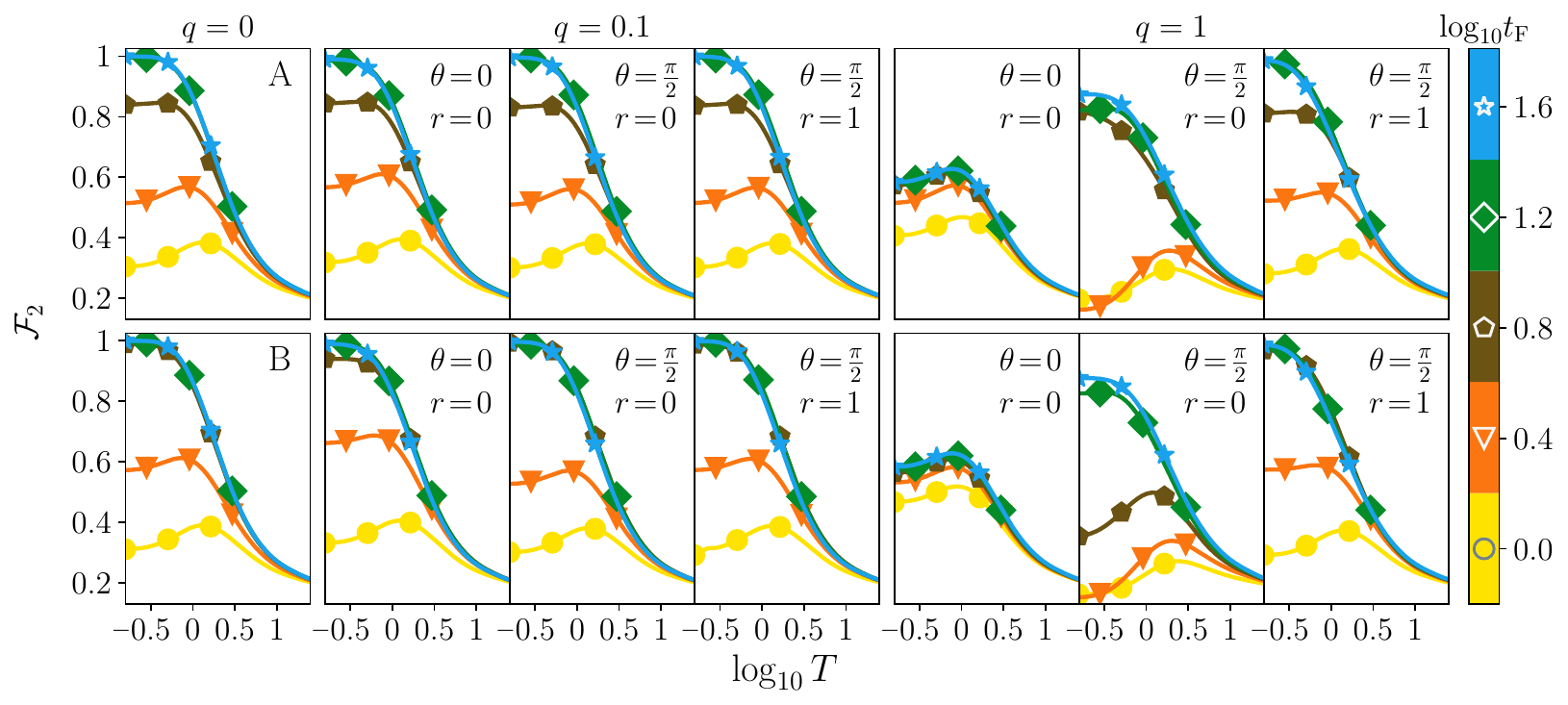}
    \caption{Same as in Fig.\,\ref{fig:First_order_fidelity} but for the path across the second-order QPT and smaller driving times.}
    \label{fig:Second_order_fidelity}
\end{figure*}

Figure~\ref{fig:Second_order_fidelity} displays the target ground-state fidelity $\mathcal{F}_2$ from Eq.\,\eqref{fidel2} for the path across the second-order QPT. 
The plots in this figure show all dependencies as in the previous case of the first-order QPT (Sec.\,\ref{se:first}), i.e., the dependence on the temperature $T$, the S-E coupling strength $q$, the angle $\theta$ of the coupling operator \eqref{Q}, the driving time~$t_{\rm F}$, the parameter ${r}$ in \eqref{counter}, and on the driving protocols A and B.
Since the minimal ground-state energy gap $\Delta_{10}$ along second-order QPT path decreases with $N$ algebraically, the driving times $t_{\rm F}$ needed to reach values of fidelity comparable to the first-order QPT path (for which the gap closes exponentially) are much shorter than in that case.
So we move from the time scale ${t_{\rm F}\in[10^{0.4},10^{3.6}]}$ used for drivings across the first-order QPT to ${t_{\rm F}\in[10^{0},10^{1.6}]}$ used for drivings across the second-order QPT.

As in the case of the first-order QPT drivings, we try to itemize the immediate observations following from Fig.\,\ref{fig:Second_order_fidelity}:
\begin{itemize}
\item[(a)] Trivially, large $t_{\rm F}$ and small $T$ imply higher fidelity than small $t_{\rm F}$ or large $T$. 
\item[(b)] For zero or weak coupling $q$ the driving protocol B still yields higher fidelity than A, but the effect is visible only for large $t_{\rm F}$.
\item[(c)] The effect of environment-induced increase of fidelity is more or less gone. Interactions with the environment start playing some role only for strong coupling (as expected) and they mostly decrease the fidelity.
\item[(d)] The high-temperature limit \eqref{limF} of fidelity is valid. 
\item[(e)] Nonmonotonous dependencies of some fidelity curves on the temperature are again observed.
\item[(f)] The counterterm in the Hamiltonian improves the fidelity for the strong coupling (compared to the ${r=0}$ case).
\item[(g)] For the strong coupling, ${q=1}$, we observe a~strong dependence of $\mathcal{F}_2$ on the coupling operator ${\hat{Q}=\hat{J}_x}$ (i.e., ${\theta = \frac{\pi}{2}}$) and~${\hat{J}_z}$ (i.e., ${\theta = 0}$).
\end{itemize}

\begin{figure}[tp]
    \includegraphics[width=\linewidth]{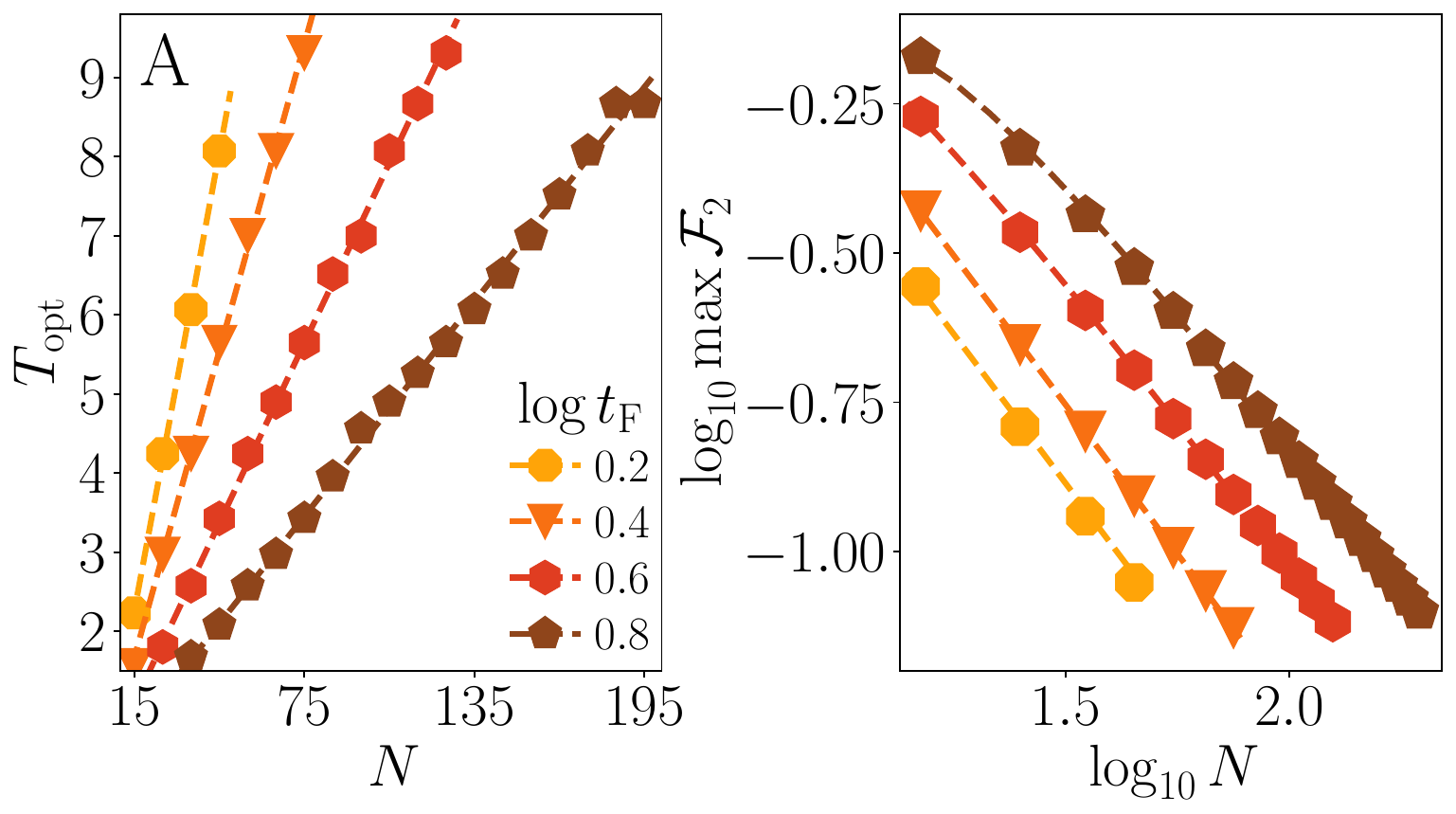}
    \caption{Same as Fig.\,\ref{fig:T_opt_andmore} but for the second-order QPT path. The dashed lines represent dependencies ${T_{\rm opt}=aN}$ and ${{\rm max}\,\mathcal{F}_2 \approx b/N - c/N^2}$ with fitted coefficients $a,b,c$.}
    \label{fig:T_opt_andmore_2ndorder}
\end{figure}

The peak of some fidelity curves, see item (e), has been subject to the same analysis as in the first-order QPT case, the result being summarized in Fig.\,\ref{fig:T_opt_andmore_2ndorder}.
We again observe the linear increase of the optimal temperature $T_{\rm opt}$ and a roughly algebraic decrease of ${\rm max}\,\mathcal{F}_2$ with the size $N$ of the qubit system.

In the case of the second-order QPT path, special attention needs to be paid to the choice of the system-environment coupling operator $\hat{Q}$ in Eq.\,\eqref{Q}.
As pointed out in item~(g), this choice makes a large difference in the values of~$\mathcal{F}_2$ obtained for the strong coupling.
This effect is due to the difference between both choices in terms of the parity conservation.
We know that for ${q=0}$ the driving from the positive-parity initial state along the ${\chi=0}$ path can excite the system only to states with positive parity.
However, for ${q\neq 0}$ the parity may be violated by system-environment interaction.
This happens for ${\theta=\frac{\pi}{2}}$, so during the drive the environment induces strong transitions to the first excited state with negative parity. 
This population is then counted in the summed final fidelity~$\mathcal{F}_2$ in Eq.\,\eqref{fidel2}, so for slow-enough driving (when transitions to higher excited states can be neglected) this fidelity may be relatively high. 

In contrast, for ${\theta=0}$ the parity is conserved even by the system-environment interaction, so the strongest transitions during the drive lead to the second excited state with positive parity, which does not contribute to~$\mathcal{F}_2$.
This explains why slow drivings in the strong-coupling columns of Fig.\,\ref{fig:Second_order_fidelity} yield higher fidelity for ${\theta=\frac{\pi}{2}}$ than for ${\theta=0}$.
In the latter case all $\mathcal{F}_2$ curves converge to a narrow band, washing out the difference between slow and fast driving.
It seems that for ${\theta=0}$ the environment-induced transitions and the transitions of the Landau-Zener type have similar effects which complement each other as the driving time~$t_{\rm F}$ varies.

\subsection{Comparison with the Lindblad method}
\label{sec:discussion}

In this section we compare the above HEOM results with simulations using the Lindblad method. 
The use of the Lindblad formalism for time-dependent Hamiltonians is discussed, e.g., in Refs.\,\cite{Alba12,Yama17,Dann18}.
In our calculations we apply the Lindblad formula
\begin{equation}\label{Lindblad}
\frac{d}{dt}\hat{\rho}_{\text{S}}(t) = -i [\hat{H}_{\text{S}}(t) + \hat{H}_{\text{L}}(t),\hat{\rho}_{\text{S}}(t)]  + \mathcal{D}_t \hat{\rho}_{\text{S}}(t)
\end{equation}
with the dissipator,
\begin{flalign}
\mathcal{D}_t \hat{\rho}_{\text{S}}(t)\! = \!\! \sum_{\epsilon}\! \frac{\Gamma(\epsilon)}{N}\,\!\!\left[\hat{S}(\epsilon)\hat{\rho}_{\text{S}}(t)\hat{S}^{\dagger}\!(\epsilon)\!-\! \frac{1}{2}\big\{ \hat{S}^{\dagger}\!(\epsilon)\hat{S}(\epsilon),\hat{\rho}_{\text{S}}(t) \big\}\right]\!.
\label{Dissipator}
\end{flalign}
Here the sum goes over all possible differences $\epsilon  = {E_m(\bold{\Lambda}(t))-E_n(\bold{\Lambda}(t))}$ between instantaneous system energies and
\begin{eqnarray}
\hat{S}(\epsilon) &=& \!\!\!\!\sum\limits_{\substack{m,n\\ E_m - E_n = \epsilon}}\!\!\!\! \bra{E_n}\hat{Q} \ket{E_m}\ket{E_n}\bra{E_m},
\\
\Gamma(\epsilon) &=& 2\big[1\!+\!n_{\rm E}(\epsilon)\big] \big[ J(\epsilon)\Theta(\epsilon)\!-\! J(-\epsilon)\Theta(-\epsilon) \big],
\label{real_part_Gamma}
\end{eqnarray}
where $\Theta$ denotes the Heaviside step function. 
The function \eqref{real_part_Gamma} is positive and satisfies ${\Gamma(0) = 4qT/\gamma}$ for the Drude-Lorentz form \eqref{J_Drude}. 
Negative or positive values of~$\epsilon$ indicate excitation or relaxation of the system, respectively, while ${\epsilon =0}$ indicates dephasing \cite{Breu02}.
The hermitian term 
\begin{equation}
\hat{H}_{\text{L}}(t) = \frac{1}{N}\! \sum_{\epsilon} \left[
\frac{c_0 \epsilon}{\gamma^2 + \epsilon^2} + \sum_{k=1}^{\infty} \frac{c_k \epsilon}{(2\pi kT)^2 + \epsilon^2}
\right]\!\hat{S}^{\dagger}\!(\epsilon)\hat{S}(\epsilon)
\end{equation}
in \eqref{Lindblad} represents the so-called Lamb shift Hamiltonian, with coefficients $c_0,c_1,\ldots$ introduced in \eqref{serc}.  

\begin{figure}[tp]
    \includegraphics[width=\linewidth]{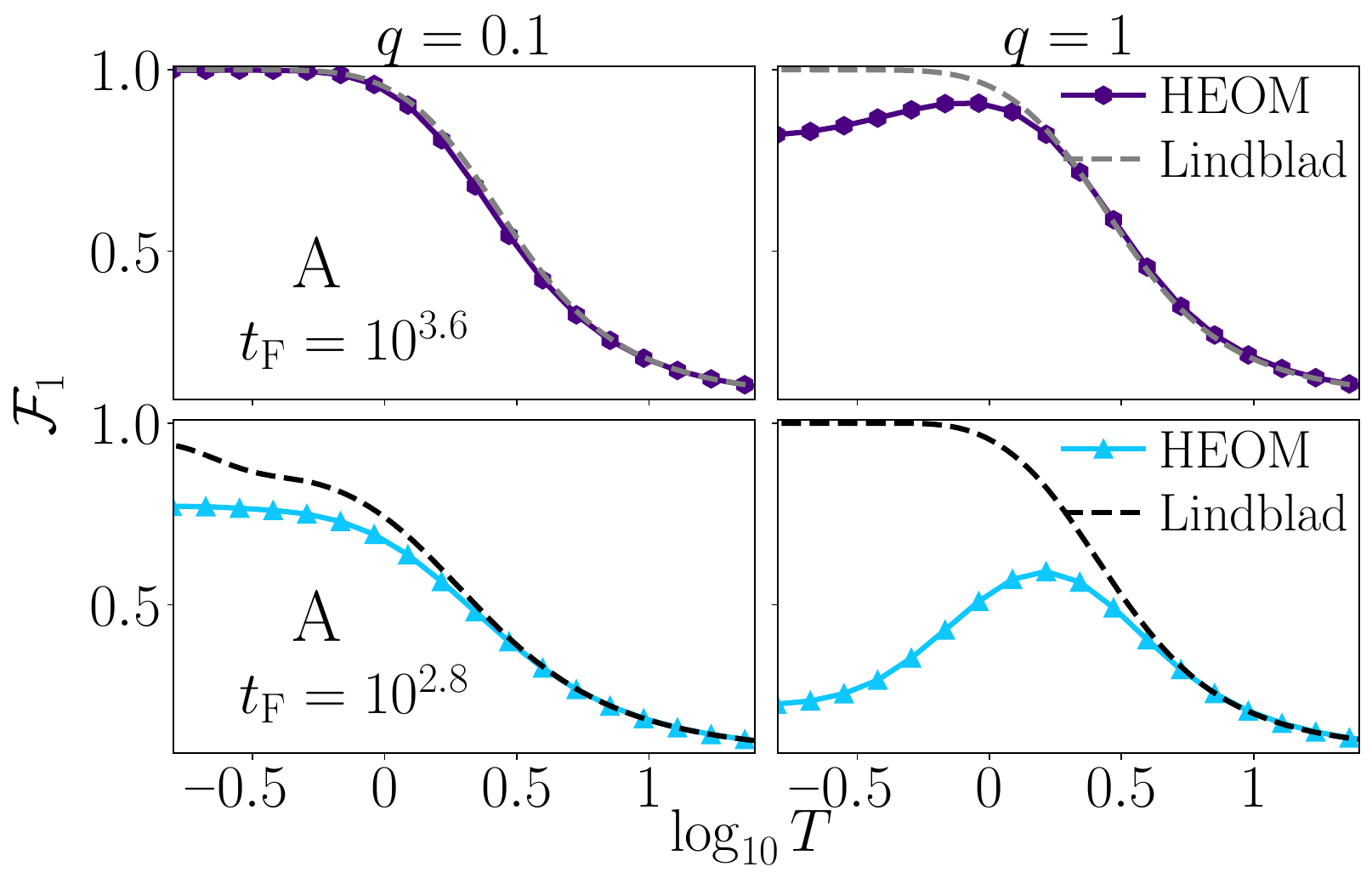}
    \caption{Fidelity for driving A across the first-order QPT (with ${\theta = 0}$ and ${r=0}$) for the ${N=10}$ qubit system obtained by the HEOM method (full curves) and by the Lindblad method according to Ref.~\cite{Yama17} (dashed curves). The driving times $t_{\rm F}$ and coupling strengths $q$ are indicated, the range of temperature is $\log_{10}T\in[-0.8,1.4]$.}
    \label{fig:Lindblad_HEOM}
\end{figure}

In the comparison of the Lindblad calculations with the HEOM results, we consider only the path across the first-order QPT with the driving protocol A (see Sec.\,\ref{se:first}).
The S-E coupling operator \eqref{Q} has ${\theta=0}$ and the counterterm in \eqref{counter} is not included (${r=0}$).
If the dynamics for ${N=10}$ qubits is modeled using the two-state approximation, the adiabatic timescale is of the order ${t = 2|d\Delta_{10}/ds|/\pi\Delta_{10}^2 \sim 10^3}$, where all terms are evaluated at the crossing~\cite{Dick13}.
The method of Ref.~\cite{Alba12,Yama17} is limited to nearly-adiabatic drivings, so only the drivings with ${t_{\rm F} \gg 10^3}$ can be expected to agree with the Lindblad calculations.

In Fig.\,\ref{fig:Lindblad_HEOM} we show the fidelity $\mathcal{F}_1$ from Eq.\,\eqref{fidel1} obtained by both the HEOM and Lindblad methods for driving times $t_{\rm F}=10^{2.8}$ and $10^{3.6}$.
We see, in accord with the above expectation, that for the larger~$t_{\rm F}$ the discrepancy between both methods affects only the drivings at lower temperatures ($T \lesssim \gamma/2\pi$, see Ref.\,\cite{Tani20}) and shows up only for the strong S-E coupling.
For this driving time, the Lindblad approach predicts the final state of the system that essentially coincides with the canonical equilibrium state $\hat{\rho}^{(\beta)}_{\rm S}(\bold{\Lambda}_{\rm F})\propto e^{-\beta \hat{H}_{\rm S}(\Lambda_{\rm F})}$.
The discord of the HEOM calculations with this prediction for the stronger coupling and low temperatures agrees with the results of Refs.\,\cite{Dijk12,Iles14,Iles16}, where deviations of the actual final state from the canonical state were also observed.
Note that a realistic approximation for the equilibrium state for all coupling strengths was derived in Refs.\,\cite{Anto23_1,Anto23_2}.
For the shorter~$t_{\rm F}$ in Fig.\,\ref{fig:Lindblad_HEOM}, the HEOM and Lindblad results differ for both coupling strengths and all temperatures below the high-$T$ limit~\eqref{limF}.

Therefore, we can conclude that non-Markovian effects, properly captured in our HEOM calculations, play in general an important role, which is not reproduced by the simpler Lindblad calculations.

\section{Conclusion}
\label{CONCLUSION}

In this paper we studied effects of thermalization and dissipation in quantum driven dynamics across finite-size precursors of QPTs of the first and second order.
Using a fully connected system of qubits coupled to an external environment at a nonzero temperature, we tested various strategies to maximize the fidelity of finite-time preparation of the pure target state associated with the entangled ground state of the qubit system at the final parameter point behind the QPT.

In agreement with our previous studies of driving in the same system without environment-induced effects~\cite{Matu23a,Matu23b} we found that the final fidelity for a fixed  finite value of the driving time can be increased when one applies the driving with constant geometric instead of planar speed.
This improvement is attributed to the suppression of exciting transitions in the QPT region due to the reduced planar speed in that region.

However, thermalization and dissipation processes can strongly modify results of the driving procedure, in some cases being able to considerably improve its fidelity.
One of the effects results from the environment-induced thermalization of the qubit system in  the course of the driving procedure, which under some circumstances increases population of the target ground state.
Another effect follows from an initial thermalization of the system at the start of the driving procedure, which for a certain optimal temperature leads to an efficient transfer of populations from excited states to the ground state in the QPT and ESQPT regions.
Both effects are present for drivings across the first- as well as second-order QPT.
The latter effect was nevertheless found to weaken with an increasing size of the system, which limits its application in driving procedures involving large numbers of qubits.

A by-product of our analysis is a comparison of the sophisticated HEOM calculations of the system-environment dynamics with much simpler and more popular Lindblad calculations.
We have demonstrated that in the present setup of driving across the finite-size QPT precursors the Lindblad calculations, which completely disregard non-Markovian effects, do not satisfactorily reproduce the more precise HEOM calculations for stronger system-environment couplings or for faster drivings. 

Finally, it needs to be stressed that all the above results were obtained in the framework of our strongly simplified model of the interacting qubit system and its environment.
Nevertheless, we consider the effects analyzed here as sufficiently robust, at least on the qualitative level, to play important roles in realistic situations, whose quantitative analysis may require additional calculations with modified model assumptions.

\section*{ACKNOWLEDGEMENTS}

We thank Jakub Dolej{\v s}{\'\i} for his help in the initial stages of this work within his diploma thesis~\cite{Dole20}.



\begin{thebibliography}{99}
\bibitem{Grov00} L.K. Grover, {\it Synthesis of Quantum Superpositions by Quantum Computation}, \href{https://journals.aps.org/prl/abstract/10.1103/PhysRevLett.85.1334}{Phys. Rev. Lett. {\bf 85}, 1334 (2000)}.
\bibitem{Ples11} M. Plesch and {\v C}. Brukner, {\it Quantum-state preparation with universal gate decompositions}, \href{https://journals.aps.org/pra/abstract/10.1103/PhysRevA.83.032302}{Phys. Rev. A {\bf 83}, 032302 (2011)}.
\bibitem{Farh00} E. Farhi, J. Goldstone, S. Gutmann, and M. Sipser, {\it Quantum Computation by Adiabatic Evolution}, \href{https://arxiv.org/abs/quant-ph/0001106}{arXiv: quant-ph/0001106 (2000)}.
\bibitem{Ahar07a} D. Aharonov and A. Ta‐Shma, {\it Adiabatic Quantum State Generation}, \href{https://doi.org/10.1137/060648829}{SIAM J. Comput. 37, 47 (2007)}.
\bibitem{Ahar07b} D. Aharonov, W. van Dam, J. Kempe, Z. Landau, S. Lloyd, and O. Regev, {\it Adiabatic Quantum Computation is Equivalent to Standard Quantum Computation}, \href{https://epubs.siam.org/doi/10.1137/S0097539705447323}{SIAM J. Comput. 37, 166 (2007).}
\bibitem{Alba18} T. Albash and D. A. Lidar,  {\it Adiabatic quantum computation}, \href{https://doi.org/10.1103/RevModPhys.90.015002}{Rev. Mod. Phys. {\bf 90}, 015002 (2018)}.
\bibitem{Niel02} M. A. Nielsen and I. Chuang, {\it Quantum Computation and Quantum Information} (Cambridge University Press, Cambridge, UK, 2010).	
\bibitem{Kato50} T. Kato,  {\it On the Adiabatic Theorem of Quantum Mechanics}, \href{https://doi.org/10.1143/JPSJ.5.435}{J. Phys. Soc. Jpn. {\bf 5}, 435 (1950)}.
\bibitem{Jans07} S. Jansen, M.-B. Ruskai, R. Seiler, {\it Bounds for the adiabatic approximation with applications to quantum computation}, \href{https://doi.org/10.1063/1.2798382}{J. Math. Phys. (N.Y.) 48, 102111 (2007)}.
\bibitem{Elga12} A. Elgart, and G.A. Hagedorn, {\it A note on the switching adiabatic theorem}, \href{https://doi.org/10.1063/1.4748968}{J. Math. Phys. (N.Y.) 53, 102202 (2012)}.
\bibitem{Demi03} M. Demirplak and S. A. Rice,  {\it Adiabatic Population Transfer with Control Fields}, \href{https://doi.org/10.1021/jp030708a}{J. Chem. Phys. {\bf 107}, 9937 (2003)}.
\bibitem{Berr09} M. V. Berry,  {\it Transitionless quantum driving}, \href{https://doi.org/10.1088/1751-8113/42/36/365303}{J. Phys. A {\bf 42}, 365303 (2009).}
\bibitem{Camp13} A. del Campo, {\it Shortcuts to Adiabaticity by Counterdiabatic Driving}, \href{https://doi.org/10.1103/PhysRevLett.111.100502}{Phys. Rev. Lett. {\bf 111}, 100502 (2013)}.
\bibitem{Guer19} D. Gu{\' e}ry-Odelin, A. Ruschhaupt, A. Kiely, E. Torrontegui, S. Mart{\' i}nez-Garaot, and J.G. Muga,  {\it Shortcuts to adiabaticity: Concepts, methods, and applications}, \href{https://doi.org/10.1103/RevModPhys.91.045001}{Rev. Mod. Phys. {\bf 91}, 045001 (2019)}.
\bibitem{Prov80} J. P. Provost and G. Vallee, {\it Riemannian structure on manifolds of quantum states}, \href{https://doi.org/10.1007/BF02193559}{Commun. Math. Phys. {\bf 76}, 289 (1980)}.
\bibitem{Anan90} J. Anandan, Y. Aharonov, {\it Geometry of quantum evolution}, \href{https://doi.org/10.1103/PhysRevLett.65.1697}{Phys. Rev. Lett. {\bf 65}, 1697 (1990)}.
\bibitem{Miya01} A. Miyake, M. Wadati, {\it Geometric strategy for the optimal quantum search}, \href{https://doi.org/10.1103/PhysRevA.64.042317}{Phys. Rev. A {\bf 64}, 042317 (2001)}.
\bibitem{Reza09} A. T. Rezakhani, W.-J. Kuo, A. Hamma, D. A. Lidar, and P. Zanardi,  {\it Quantum Adiabatic Brachistochrone}, \href{https://doi.org/10.1103/PhysRevLett.103.080502}{Phys. Rev. Lett. {\bf 103}, 080502 (2009)}.
\bibitem{Kolo17} M. Kolodrubetz, D. Sels, P. Mehta, and A. Polkovnikov, {\it Geometry and non-adiabatic response in quantum and classical systems}, \href{https://doi.org/10.1016/j.physrep.2017.07.001}{Phys. Rep. {\bf 697}, 1 (2017)}.
\bibitem{Buko19} M. Bukov, D. Sels, and A. Polkovnikov, {\it Geometric Speed Limit of Accessible Many-Body State Preparation}, \href{https://doi.org/10.1103/PhysRevX.9.011034}{Phys. Rev. X {\bf 9}, 011034 (2019)}.
\bibitem{Chil02} A.M. Childs, E. Deotto, E. Farhi, J.Goldstone, S. Gutmann, and A.J. Landahl, {\it Quantum search by measurement}, \href{https://doi.org/10.1103/PhysRevA.66.032314}{Phys. Rev. A {\bf 66}, 032314 (2002)}. 
\bibitem{Roa06} L. Roa, A. Delgado, M.L. Ladr{\'o}nde Guevara, and A.B. Klimov, {\it Measurement-driven quantum evolution}, \href{https://doi.org/10.1103/PhysRevA.73.012322}{Phys. Rev. A {\bf 73}, 012322 (2006)}. 
\bibitem{Haco18} S. Hacohen-Gourgy, L.P. Garc{\'\i}a-Pintos, L.S. Martin, J. Dressel, and I. Siddiqi, {\it Incoherent Qubit Control Using the Quantum Zeno Effect}, \href{https://doi.org/10.1103/PhysRevLett.120.020505}{Phys. Rev. Lett. {\bf 120}, 020505 (2018)}.
\bibitem{Cejn23} P. Cejnar, P. Str{\' a}nsk{\' y}, J. St{\v r}ele{\v c}ek, and F. Matus, {\it Decoherence-assisted quantum driving}, \href{https://doi.org/10.1103/PhysRevA.107.L030603}{Phys. Rev. A {\bf 107}, L030603 (2023)}.
\bibitem{Matu23a} F. Matus, J. St\ifmmode \check{r}\else \v{r}\fi{}ele\ifmmode \check{c}\else \v{c}\fi{}ek, P. Str{\' a}nsk{\' y}, and P. Cejnar, {\it Search for optimal driving in finite quantum systems with precursors of criticality}, \href{https://doi.org/10.1103/PhysRevA.107.012216}{Phys. Rev. A {\bf 107}, 012216 (2023)}.
\bibitem{Matu23b} F. Matus, J. St{\v r}ele{\v c}ek, and P. Cejnar, {\it Analytic approach to the Landau–Zener problem in bounded parameter space}, \href{https://doi.org/10.1088/1751-8121/accf4f}{J. Phys. A: Math. Theor. {\bf 56}, 235303 (2023)}.
\bibitem{Lipk65} H. J. Lipkin, N. Meshkov, and A. J. Glick,  {\it Validity of many-body approximation methods for a solvable model: (I). Exact solutions and perturbation theory}, \href{https://doi.org/10.1016/0029-5582(65)90862-X}{Nucl. Phys. {\bf 62}, 188 (1965)}. 
\bibitem{Gilm78} R. Gilmore and D. H. Feng, {\it Phase transitions in nuclear matter described by pseudospin Hamiltonians}, \href{https://doi.org/10.1016/0375-9474(78)90260-9}{Nucl. Phys. A {\bf 301}, 189 (1978)}.
\bibitem{Orus08} R. Or{\'u}s, S. Dusuel, and J. Vidal, {\it Equivalence of Critical Scaling Laws for Many-Body Entanglement in the Lipkin-Meshkov-Glick Model}, \href{https://doi.org/10.1103/PhysRevLett.101.025701}{Phys. Rev. Lett. {\bf 101}, 025701 (2008)}.
\bibitem{Zibo10} T. Zibold, E. Nicklas, C. Gross, and M.K. Oberthaler, {\it Classical Bifurcation at the Transition from Rabi to Josephson Dynamics}, \href{https://doi.org/10.1103/PhysRevLett.105.204101}{Phys. Rev. Lett. {\bf 105}, 204101 (2010)}.
\bibitem{Puri17} S. Puri, C. K. Andersen, A. L. Grimsmo, and A. Blais, {\it Quantum annealing with all-to-all connected nonlinear oscillators}, \href{https://doi.org/10.1038/ncomms15785}{Nat. Commun. {\bf 8}, 15785 (2017)}. 
\bibitem{Cerv21} M. J. Cervia, A. B. Balantekin, S. N. Coppersmith, C. W. Johnson, P. J. Love, C. Poole, K. Robbins, and M. Saffman, {\it Lipkin model on a quantum computer}, \href{https://doi.org/10.1103/PhysRevC.104.024305}{Phys. Rev. C {\bf 104}, 024305 (2021)}.
\bibitem{Sach99} S. Sachdev, {\it Quantum Phase Transitions} (Cambridge University Press, Cambridge, UK, 1999).
\bibitem{Lato04} J. I. Latorre and R. Or{\'u}s, {\it Adiabatic quantum computation and quantum phase transitions}, \href{https://journals.aps.org/pra/abstract/10.1103/PhysRevA.69.062302}{
    Phys. Rev. A {\bf 69}, 062302 (2004)}.
\bibitem{Zure05} W.H. Zurek, U. Dorner, and P. Zoller, {\it Dynamics of a Quantum Phase Transition}, \href{https://doi.org/10.1103/PhysRevLett.95.105701}{Phys. Rev. Lett. {\bf 95}, 105701 (2005)}.
\bibitem{Schu06} R. Sch{\"u}tzhold and G. Schaller, {\it Adiabatic quantum algorithms as quantum phase transitions: First versus second order}, \href{https://doi.org/10.1103/PhysRevA.74.060304}{Phys. Rev. A {\bf 74}, 060304(R) (2006)}.
\bibitem{Cejn06} P. Cejnar, M. Macek, S. Heinze, J. Jolie, and J. Dobe{\v s}, {\it Monodromy and excited-state quantum phase transitions in integrable systems: collective vibrations of nuclei}, \href{https://iopscience.iop.org/article/10.1088/0305-4470/39/31/L01}{J. Phys. A: Math. Gen. {\bf 39}, L515 (2006)}.
\bibitem{Ribe07} P. Ribeiro, J. Vidal, and R. Mosseri, {\it Thermodynamical Limit of the Lipkin-Meshkov-Glick Model}, \href{https://doi.org/10.1103/PhysRevLett.99.050402}{Phys. Rev. Lett. {\bf 99}, 050402 (2007)}.
\bibitem{Capr08} M. Caprio, P. Cejnar, and F. Iachello, {\it Excited state quantum phase transitions in many-body systems}, \href{https://doi.org/10.1016/j.aop.2007.06.011}{Ann. Phys. {\bf 323}, 1106 (2008)}.
\bibitem{Cejn08} P. Cejnar, P. Str{\'a}nsk{\'y}, {\it Impact of quantum phase transitions on excited-level dynamics},
\href{https://doi.org/10.1103/PhysRevE.78.031130}{Phys. Rev. E {\bf 78}, 031130 (2008)}.
\bibitem{Pere09} P. P{\' e}rez-Fern{\' a}ndez, A. Rela\~{n}o, J. M. Arias, J. Dukelsky, and J. E. Garc{\' i}a-Ramos, {\it Decoherence due to an excited-state quantum phase transition in a two-level boson model}, \href{https://doi.org/10.1103/PhysRevA.80.032111}{Phys. Rev. A {\bf 80}, 032111 (2009)}.	
\bibitem{Sant16} L. F. Santos, M. T{\' a}vora and F. P{\' e}rez-Bernal, {\it Excited-state quantum phase transitions in many-body systems with infinite-range interaction: Localization, dynamics, and bifurcation}, \href{https://doi.org/10.1103/PhysRevA.94.012113}{Phys. Rev. A {\bf 94}, 012113 (2016)}.
\bibitem{Kopy17} W. Kopylov, G. Schaller, and T. Brandes, {\it Nonadiabatic dynamics of the excited states for the Lipkin-Meshkov-Glick model}, \href{https://journals.aps.org/pre/abstract/10.1103/PhysRevE.96.012153}{Phys. Rev. E {\bf 96}, 012153 (2017)}.
\bibitem{Cejn21} P. Cejnar, P. Stránský, M. Macek, M. Kloc, {\it Excited-state quantum phase transitions}, \href{https://iopscience.iop.org/article/10.1088/1751-8121/abdfe8}{J. Phys. A: Math. Theo. {\bf 54}, 133001 (2021)}.
\bibitem{Dick13} N. Dickson, M. Johnson, M. Amin {\it et al}, {\it Thermally assisted quantum annealing of a 16-qubit problem}, \href{https://doi.org/10.1038/ncomms2920}{Nat. Commun. {\bf 4}, 1903 (2013)}.
\bibitem{Alba17} T. Albash, V. Martin-Mayor and I. Hen, {\it Temperature Scaling Law for Quantum Annealing Optimizers}, \href{https://doi.org/10.1103/PhysRevLett.119.110502}{Phys. Rev. Lett. {\bf 119}, 110502 (2017)}.
\bibitem{Breu02} H.-P. Breuer and F. Petruccione, {\it The Theory of Open Quantum Systems} (Clarendon Press, Oxford, 2002).
\bibitem{Weis12} U. Weiss, {\it Quantum Dissipative Systems} (World Scientific, 4th ed., Singapore, 2012).
\bibitem{Schl07} M. Schlosshauer, {\it Decoherence and the Quantum-to-classical transition} (Springer, 1st ed., Berlin, 2010).
\bibitem{Alba12} T. Albash, S. Boixo, D. A. Lidar, and P. Zanardi, {\it Quantum adiabatic Markovian master equations}, \href{https://doi.org/10.1088/1367-2630/14/12/123016}{New J. Phys. {\bf 14}, 123016 (2012)}.	
\bibitem{Yama17} M. Yamaguchi, T. Yuge and T. Ogawa, {\it Markovian quantum master equation beyond adiabatic regime}, \href{https://doi.org/10.1103/PhysRevE.95.012136}{Phys. Rev. E {\bf 95}, 012136 (2017)}.	
\bibitem{Dann18} R. Dann, A. Levy, and R. Kosloff, {\it Time-dependent Markovian quantum master equation}, \href{https://doi.org/10.1103/PhysRevA.98.052129}{Phys. Rev. A {\bf 98}, 052129 (2018)}.	
\bibitem{Ishi05} A. Ishizaki1 and Y., Tanimura, {\it Quantum Dynamics of System Strongly Coupled to Low-Temperature Colored Noise Bath: Reduced Hierarchy Equations Approach}, \href{https://doi.org/10.1143/JPSJ.74.3131}{J. Phys. Soc. Jpn. {\bf 74},  3131-3134 (2005)}. 
\bibitem{Dijk12} A. G. Dijkstra and Y. Tanimura, {\it System Bath Correlations and the Nonlinear Response of Qubits}, \href{https://doi.org/10.1143/JPSJ.81.063301}{J. Phys. Soc. Jpn. {\bf 81}, 063301 (2012)}. 
\bibitem{Iles14} J. Iles-Smith, N. Lambert, and A. Nazir, {\it Environmental dynamics, correlations, and the emergence of noncanonical equilibrium states in open quantum systems}, \href{https://doi.org/10.1103/PhysRevA.90.032114}{Phys. Rev. A {\bf 90}, 032114 (2014)}. 
\bibitem{Iles16} J. Iles-Smith, A.G. Dijkstra, N. Lambert, and A. Nazir, {\it Energy transfer in structured and unstructured environments: Master equations beyond the Born-Markov approximations}, \href{https://doi.org/10.1063/1.4940218}{J. Chem. Phys. {\bf 144}, 044110 (2016)}.
\bibitem{Abbo20} J. W. Abbott, {\it Quantum Dynamics of Bath Influenced Excitonic Energy Transfer in Photosynthetic Pigment-Protein Complexes Creators}, \href{https://zenodo.org/records/7229807}{Master Thesis (2020)}
\bibitem{Anto23_1} N. Anto-Sztrikacs, A. Nazir, and D. Segal, {\it Effective Hamiltonian theory of open quantum systems at strong coupling}, \href{https://doi.org/10.1103/PRXQuantum.4.020307}{PRX Quantum {\bf 4}, 020307 (2023)}
\bibitem{Anto23_2} N. Anto-Sztrikacs, B. Min, M. Brenes and D. Segal, {\it Effective Hamiltonian theory: An approximation to the equilibrium state of open quantum systems}, \href{https://doi.org/10.1103/PhysRevB.108.115437}{Phys. Rev. B  {\bf 108}, 115437 (2023)}
\bibitem{Tani89} Y. Tanimura and R. Kubo, {\it Time Evolution of a Quantum System in Contact with a Nearly Gaussian-Markoffian Noise Bath}, \href{https://doi.org/10.1143/JPSJ.58.101}{J. Phys. Soc. Jpn. {\bf 58}, 101 (1989)}.
\bibitem{Tani06} Y. Tanimura, {\it Stochastic Liouville, Langevin, Fokker–Planck, and Master Equation Approaches to Quantum Dissipative Systems}, \href{https://doi.org/10.1143/JPSJ.75.082001}{J. Phys. Soc. Jpn. {\bf 75},  082001 (2006)}.
\bibitem{Fruch16} A. Fruchtman, N. Lambert, and E.M. Gauger, {\it When do perturbative approaches accurately capture the dynamics of complex quantum systems?}, \href{https://doi.org/10.1038/srep28204}{Sci. Rep. {\bf 6}, 28204 (2016)}.
\bibitem{Tani20} Y. Tanimura, {\it Numerically “exact” approach to open quantum dynamics: The hierarchical equations of motion (HEOM)}, \href{https://doi.org/10.1063/5.0011599}{J. Chem. Phys. {\bf 153}, 020901 (2020)}.
\bibitem{Lamb23} N. Lambert, T. Raheja, S. Cross, P. Menczel, S. Ahmed, A. Pitchford, D. Burgarth, and F. Nori, {\it QuTiP-BoFiN: A bosonic and fermionic numerical hierarchical-equations-of-motion library with applications in light-harvesting, quantum control, and single-molecule electronics}, \href{https://doi.org/10.1103/PhysRevResearch.5.013181}{Phys. Rev. Research {\bf 5}, 013181 (2023)}.
\bibitem{Feyn63} R. P. Feynman and F. L. Vernon Jr., {\it The theory of a general quantum system interacting with a linear dissipative system}, \href{https://doi.org/10.1016/0003-4916(63)90068-X}{Ann. Phys. (USA) {\bf 24}, 118 (1963)}. 
\bibitem{Cald81} A. O. Caldeira and A. J. Leggett, {\it Influence of Dissipation on Quantum Tunneling in Macroscopic Systems}, \href{https://doi.org/10.1103/PhysRevLett.46.211}{Phys. Rev. Lett. {\bf 46}, 211 (1981)}.
\bibitem{Cald83} A. O. Caldeira and A. J. Leggett, {\it Quantum tunnelling in a dissipative system}, \href{https://doi.org/10.1016/0003-4916(83)90202-6}{Ann. Phys. (NY) {\bf 149}, 374 (1983)}.
\bibitem{Xu09} J. Xu, R.-X. Xu, and Y.J. Yan, {\it Exact quantum dissipative dynamics under external time-dependent driving fields}, \href{https://iopscience.iop.org/article/10.1088/1367-2630/11/10/105037}{New J. Phys. {\bf 11}, 105037 (2009)}.
\bibitem{Ma12} J. Ma, Z. Sun, X. Wang, and F. Nori, {\it Entanglement dynamics of two qubits in a common bath}, \href{https://doi.org/10.1103/PhysRevA.85.062323}{Phys. Rev. A {\bf 85}, 062323 (2012)} .
\bibitem{Ishi09} A. Ishizaki and G. R. Fleming, {\it Theoretical examination of quantum coherence in a photosynthetic system at physiological temperature}, \href{https://doi.org/10.1073/pnas.0908989106}{PNAS {\bf 106}, 17255 (2009)}.
\bibitem{Chen09} L. Chen, R. Zheng, Q. Shi, and Y. Yan, {\it Optical line shapes of molecular aggregates: Hierarchical equations of motion method}, \href{https://doi.org/10.1063/1.3213013}{J. Chem. Phys. {\bf 131}, 094502 (2009)}.
\bibitem{Stru11} J. Str{\"u}mpfer and K. Schulten, {\it The effect of correlated bath fluctuations on exciton transfer}, \href{https://doi.org/10.1063/1.3557042}{J. Chem. Phys. {\bf 134}, 095102 (2011)}.
\bibitem{Lamb19} N. Lambert, S. Ahmed, M. Cirio, and F. Nori, {\it Modelling the ultra-strongly coupled spin-boson model with unphysical modes}, \href{https://doi.org/10.1038/s41467-019-11656-1}{Nat. Commun. {\bf 10}, 3721 (2019)}.
\bibitem{Cejn16} P. Cejnar and P. Str{\'a}nsk{\'y}, {\it Quantum phase transitions in the collective degrees of freedom: nuclei and other many-body systems}, \href{https://iopscience.iop.org/article/10.1088/0031-8949/91/8/083006}{Phys. Scr. {\bf 91}, 083006 (2016)}.
\bibitem{Strel24} J. St{\v r}ele{\v c}ek and P. Cejnar, {\it Quantum geometry in many-body systems with precursors of criticality}, \href{https://doi.org/10.48550/arXiv.2411.03967}{arXiv: quant-ph/2408.00635 (2024)}.
\bibitem{Rzaz75} K. Rza\.{z}ewski, K. W{\'o}dkiewicz, and W. \.{Z}akowicz, {\it Phase Transitions, Two-Level Atoms, and the $A^2$ Term}, \href{https://doi.org/10.1103/PhysRevLett.35.432}{Phys. Rev. Lett. {\bf 35}, 432 (1975)}.
\bibitem{Hepp73} K. Hepp and E. H. Lieb, {\it On the superradiant phase transition for molecules in a quantized radiation field: the Dicke maser model},
\href{https://doi.org/10.1016/0003-4916(73)90039-0}{Ann. Phys. {\bf 76}, 360 (1973)}.
\bibitem{Wang73} Y. K. Wang and F. T. Hioe, {\it Phase Transition in the Dicke Model of Superradiance}, \href{https://doi.org/10.1103/PhysRevA.7.831}{Phys. Rev. A {\bf 7}, 831 (1973)}.
\bibitem{Emary03} C. Emary and T. Brandes, {\it Quantum Chaos Triggered by Precursors of a Quantum Phase Transition: The Dicke Model}, \href{https://doi.org/10.1103/PhysRevLett.90.044101}{Phys. Rev. Lett. {\bf 90}, 044101 (2003)}.
\bibitem{Baum10} K. Baumann, C. Guerlin, F. Brennecke and T. Esslinger, {\it Dicke quantum phase transition with a superfluid gas in an optical cavity}, \href{https://doi.org/10.1038/nature09009}{Nature {\bf 464}, 1301 (2010)}.
\bibitem{Nata10} P. Nataf and C. Ciuti, {\it No-go theorem for superradiant quantum phase transitions in cavity QED and counter-example in circuit QED}, \href{https://doi.org/10.1038/ncomms1069}{Nat Commun {\bf 1}, 72 (2010)}.
\bibitem{Kopyl19} W. Kopylov and G. Schaller, {\it Polaron-transformed dissipative Lipkin-Meshkov-Glick model}, \href{https://doi.org/10.1103/PhysRevA.100.063815}{Phys. Rev. A {\bf 100}, 063815 (2019)}.
\bibitem{Rola02} J. Roland and N. J. Cerf, {\it Quantum search by local adiabatic evolution}, \href{https://link.aps.org/doi/10.1103/PhysRevA.65.042308}{Phys. Rev. A {\bf 65}, 042308 (2002)}.
\bibitem{Wubs06} M. Wubs, K. Saito, S. Kohler, P. H{\"a}nggi and Y. Kayanuma, {\it Gauging a Quantum Heat Bath with Dissipative Landau-Zener Transitions}, \href{https://doi.org/10.1103/PhysRevLett.97.200404}{Phys. Rev. Lett. {\bf 97}, 200404 (2006)}.
\bibitem{Dole20} J. Dolej{\v s}{\'\i}, Diploma Thesis (Faculty of Mathematics and Physics, Charles University, 2020).
\end{thebibliography}
\end{document}